\documentclass[preprintnumbers,superscriptaddress,nofootinbib,aps,prd,floatfix]{revtex4}

\pdfoutput=1
\usepackage{graphicx}
\usepackage{latexsym}
\usepackage{amsfonts}
\usepackage{amssymb}
\usepackage{amsmath}
\usepackage{slashed}
\usepackage{feynmp}
\usepackage{hyperref}
\usepackage{url}
\usepackage{color}
\usepackage{cancel}
\usepackage{multirow,array}
\usepackage{float}
\usepackage{subfigure}

\newcommand{\gaia}{\textsc{Gaia}}
\newcommand{\dama}{\textsc{DAMA/Libra}}
\newcommand{\xenon}{\textsc{Xenon1T}}
\newcommand{\cdms}{\textsc{CDMSlite}}
\newcommand{\pico}{\textsc{PICO-60}}
\newcommand{\cosine}{\textsc{COSINE-100}}
\newcommand{\anais}{\textsc{ANAIS-112}} 
\begin{document}

\date{\today}

\title{Direct Detection Anomalies in light of {\it Gaia} Data}

\author{Matthew R.~Buckley}
\affiliation{Department of Physics and Astronomy, Rutgers University, Piscataway, NJ 08854, USA}

\author{Gopolang Mohlabeng}
\affiliation{Physics Department, Brookhaven National Laboratory, Upton, New York 11973, USA}

\author{Christopher W.~Murphy}
\affiliation{Physics Department, Brookhaven National Laboratory, Upton, New York 11973, USA}
\affiliation{Insight Data Science, San Francisco, California 94107, USA }

\begin{abstract}
Measurements from the \gaia~satellite have greatly increased our knowledge of the dark matter velocity distributions in the Solar neighborhood. There is evidence for multiple cold structures nearby, including a high-velocity stream counterrotating relative to the Sun. This stream could significantly alter the spectrum of recoil energies and increase the annual modulation of dark matter in direct detection experiments such as \dama.  We reanalyze the experimental limits from \xenon, \cdms, \pico~and \cosine, and compare them to the results of the \dama~experiment. While we find that this new component of the dark matter velocity distribution can greatly improve the fit to the \dama~data, both spin-independent and spin-dependent interpretations of the \dama~signal with elastic and inelastic scattering continue to be ruled out by the null results of other experiments, in particular \xenon.
\end{abstract}

\maketitle


\section{Introduction}

There is strong gravitational evidence that 26.5\% of the Universe's energy budget and the majority of the mass within galaxy clusters is composed of dark matter; an unknown, invisible, non-relativistic material. No particle in the Standard Model (SM) has the appropriate properties to be dark matter, thus evidence of dark matter is evidence of new physics. If dark matter consists of a particle with a mass $\gtrsim $~GeV and has small but non-zero couplings to the SM baryons (as suggested by the thermal relic freeze-out scenario), then direct detection experiments can be a powerful probe of the dark sector. 

Direct detection searches for a dark matter particle passing through a low-background detector undergoing a nuclear recoil with one of the detector's atoms. This recoil can then be seen in the experiment, assuming that the recoil momentum is above the detector threshold. The number of events in a given direct detection experiment is then set by a confluence of factors: the parameters in the dark matter theory space (e.g., mass, scattering mechanism, and cross section), the detector energy threshold and efficiency function, and the astrophysical distribution of dark matter (the local density and local velocity distribution). From a particle theorist's point of view, these latter parameters are confounding variables which prevent a straightforward mapping from the experimental results to exclusion or discovery within the theory space. 

In particular, if dark matter near the Sun is moving with velocities significantly different from the baseline expectations, then the relative sensitivity between different experiments can also be drastically altered. Indeed, as has been well-studied, dark matter only $N$-body simulations \cite{Vogelsberger:2008qb,Kuhlen:2009vh,Ling:2009eh,Lisanti:2011as,Butsky:2015pya,Bozorgnia:2016ogo}, find large deviations from the Maxwell-Boltzmann Standard Halo Model (SHM). While baryonic effects are found to bring the velocity distribution more in line with the SHM \cite{Bozorgnia:2016ogo,Kelso:2016qqj,Sloane:2016kyi,Bozorgnia:2017brl}, there is also some variation of velocity distributions between simulated galaxies, and in most examples the high-velocity tail differs significantly from the SHM.
Since the Milky Way halo was constructed through the hierarchical merger of smaller subhalos \cite{1978MNRAS.183..341W}, we should expect streams and tidal debris \cite{Lisanti:2011as} of dark matter, giving additional velocity structure on top of the smooth halo. 
These results have significant implications for dark matter direct detection, even if the stream contributes only a small component of the dark matter density \cite{Savage:2006qr,Kuhlen:2012fz}.

Given this background, the results of the \gaia~mission \cite{Prusti:2016bjo} are of timely interest. \gaia~measures the position and  motion of the nearest $\sim 1.4$ billion stars, mostly within $\sim 2$~kpc of the Sun's location. Combined with metallicity information, this enormous sample allows for kinematic studies of the stellar halo with a resolution and coverage not possible before. 
Though the stellar halo is not expected to trace the full phase-space structure of dark matter \cite{Bozorgnia:2018pfa} -- missing in particular that component accreted from non-luminous substructure \cite{Necib:2018igl} -- the low and intermediate-metallicity stars in the halo are expected to trace the earliest-accreted component of the dark matter \cite{Herzog-Arbeitman:2017fte, Necib:2018igl}. Hence \gaia~can infer a great deal about the dark matter velocity distribution in the Solar neighborhood. 
Several analyses \cite{Necib:2018iwb,Necib:2018igl} find that the smooth halo component of dark matter is suppressed at high-velocities (as expected from N-body simulations \cite{Vogelsberger:2008qb,Kuhlen:2009vh,Ling:2009eh,Lisanti:2011as,Butsky:2015pya}), leading to weaker direct detection sensitivity for low mass dark matter as compared to the predictions of the SHM, though Ref.~\cite{Evans:2018bqy} finds a velocity profile that is instead shifted towards higher velocities and stronger limits, behavior that was also found in simulation by Ref.~\cite{Bozorgnia:2016ogo}.
Perhaps more surprisingly, there is evidence that the Sun is in the path of a number of high velocity dark matter streams \cite{Myeong:2017skt}. One such stream (the S1 stream)  is counter-rotating relative to the Sun's motion through the Galaxy, increasing the relative motion of the dark matter within the stream and the Earth.  This stream is consistent with being part of a remnant of a $\sim 2\times 10^{10}\,M_\odot$ dwarf galaxy, which was tidally stripped by the Milky Way several billion years ago. While the dark matter density of S1 has not yet been measured, it could be a significant component locally, perhaps ${\cal O}(10\%)$ \cite{OHare:2018trr}.

The impact of both the \gaia-derived halo velocity distribution and that of the S1 stream on the future direct detection limits have been studied in prior literature \cite{OHare:2018trr,Necib:2018iwb}. However, until this work, their effect on existing limits and potential signals have not been fully calculated. Of particular interest is the effect of S1 on direct detection experiments that search for an annual {\it modulation} of scattering events, rather than the overall rate averaged over a year. As the Earth moves around the Sun, the fraction of dark matter particles with a relative velocity capable of surpassing the detector threshold increases when the Earth's motion is into the local dark matter ``wind,'' and decreases as the Earth moves with the flow of dark matter. As a result, the direct detection rate should modulate over a year.

As the S1 stream is itself fast-moving and counter-rotating to the Sun's motion, it could induce a very large annual modulation -- in some cases nearly an order of magnitude more than one would expect from the dark matter halo itself. As the stream is kinematically cold, the velocity profile is narrow, leading to a sharp peak in the nuclear recoil spectrum, as would be seen by a direct detection experiment (at a recoil energy set by a combination of the dark matter mass and the target nuclei). Additionally, the S1 stream happens to be oriented in such a way that the modulation peak occurs on a day in early June, near the day one would predict from the non-rotating halo. 

As is well-known, the \dama~experiment \cite{Bernabei:2008yh} observes an annual modulation in scattering events with a peak date in June \cite{Bernabei:2018yyw}. 
This signal is also compatible with the initial modulation analysis of the \cosine~\cite{Adhikari:2019off} and \anais~\cite{Amare:2019jul} experiments, though the results have nearly equal preference for the null (i.e., no dark matter signal) hypothesis.
Assuming the SHM or a smooth halo profile derived from simulation, it is difficult to interpret the \dama~signal as dark matter, given the negative results from other experiments, for both spin-independent (SI) and spin-dependent (SD) interactions \cite{Savage:2008er,Savage:2010tg,McCabe:2011sr,Catena:2016hoj,Baum:2018ekm,Kang:2018qvz}.  Indeed, the energy distribution of the \dama~signal itself is difficult to reconcile with a dark matter signal under the assumptions of the SHM \cite{Baum:2018ekm}.
However, in light of the \gaia~data and the S1 stream we should revisit this conclusion, as these could greatly increase the annual modulation signal without as significant a change in the limits set by other experiments, as well as alter the observed recoil spectrum in a way that might better fit the data.

Given this motivation, using the velocity distributions extracted from \gaia~data, we reanalyze the direct detection limits from the \dama~experiment along with the experiments that set the strongest current limits for spin-dependent and spin-independent scattering: \xenon~\cite{Aprile:2018dbl}, \cdms~\cite{Agnese:2017jvy}, \pico~\cite{Amole:2017dex}, and the \cosine~rate measurement \cite{Adhikari2018}, considering a range of possibilities for the couplings of dark matter to protons and neutrons. 
We show that  \gaia-derived halo models mildly weaken the bounds on dark matter interactions with baryons for lower mass dark matter when compared with the predictions from the SHM. Addition of the S1 stream can improve the statistical fit of the \dama~signal to a dark matter interpretation (compared to the fit to a dark matter velocity distribution without a stream), and shifts the best-fit region to lower masses and cross sections. Despite these significant changes, we find that for all tested scenarios of elastic and inelastic scattering (assuming either SD or SI coupling) the best-fit regions of the \dama~signal continue to be excluded by null results of other experiments (most importantly, \xenon), unless the S1 stream is the dominant contributor to the local dark matter density ($\gtrsim 80\%$).

After a review of dark matter scattering physics in Section~\ref{sec:scattering}, in Section~\ref{sec:distribution} we examine the local velocity distribution as determined by the \gaia~satellite in the context of dark matter direct detection experiments, including both the background halo distribution and the S1 stream. We then examine the impact of these velocity distributions on the existing experimental limits of \xenon, \cdms, \cosine, and \pico, and the best-fit regions of \dama, for both SD and SI couplings, varying degrees of isospin violation, and varying S1 dark matter densities. Details of our recasting of experimental results are described in Appendix~\ref{app:recast}. 

\section{Review of Dark Matter Direct Detection \label{sec:scattering}}
The differential rate $dR/dE_R$ for dark matter scattering off a target nucleus of atomic mass $M$ in a particular direct detection experiment is given by 
\begin{equation}
\frac{dR}{dE_R} = \frac{\rho_0}{2\mu^2m_\chi} \sigma_N F^{2}(E_R) \int_{v_{\rm min}(E_R)}^\infty v^2 dv \int d\Omega \frac{\tilde{f}(v)}{v}, \label{eq:dRdER}
\end{equation}
expressed in counts/day/kg/keV with $E_R$ the nuclear recoil energy.
The quantity $\rho_0$ is the local dark matter density (to which we assign a canonical value\footnote{This canonical choice is likely an underestimate of the local density, which recent surveys estimate to be $0.4-0.7$~GeV/cm$^3$ \cite{Smith:2011fs,Bienayme:2014kva,Piffl:2014mfa,Sivertsson:2017rkp,2018A&A...615A..99H,Read:2014qva,Buch:2018qdr}. This would result in a straightforward rescaling of our results equally for all experiments.} of $0.3$~GeV/cm$^3$), $\mu = Mm_\chi/(m_\chi +M)$ is the nucleus-dark matter reduced mass, and $F(q)$ is a form-factor which depends on the transferred momentum $q = \sqrt{2 M E_R}$.
For spin-independent scattering we adopt the Helm form factor \cite{Helm:1956zz}
\begin{equation}
F(q) = \frac{3 (\sin{q r} - q r \cos{q r})}{(q r)^{3}} e^{\frac{(q s)^{2}}{2}},
\end{equation}
where (following Ref.~\cite{Lewin:1995rx}), $r = \sqrt{c^{2} + 7/3 ~\pi^{2} a^{2} - 5 s^{2}}$ is the nuclear radius, $c = 1.23 A^{1/3} - 0.6$,  $a = 0.52$, $s = 0.9$, with $A$ the atomic mass of a specific nuclear target. 
The spin-dependent form-factor goes as
\begin{equation}
F^{2}(q) = \frac{S(q)}{S(0)},
\end{equation}
where $S(q) = a_{0}^{2} S_{00} (q) + a_{0} a_{1} S_{01}(q) + a_{1}^{2} S_{11}(q)$ \cite{Jungman:1995df, Lewin:1995rx, Klos:2013rwa, Servant:2002hb} and $S(0)$ is the normalization of the structure functions in the zero momentum transfer limit. The $S_{ij}$ contain information on the protons, neutrons, and their interference. For the different target nuclei we use the nuclear shell model fits derived in the appendix of Ref.~\cite{Klos:2013rwa} and for the nucleon coefficients $a_{i}$, we use the values in Refs.~\cite{Belanger:2008sj, Backovic:2015cra}.\\

The quantity $\tilde{f}(v)$ in Eq.~\eqref{eq:dRdER} is the lab frame dark matter velocity distribution and is integrated from the detector dependent velocity $v_{\rm min}(E_R)$ to the maximum velocity where the density distribution has support. This is set by the escape velocity of the Galaxy, $v_{\rm esc}$.
The velocity $v_{\rm min}$ is the minimum dark matter velocity capable of inducing a nuclear scattering event with recoil energy $E_R$. For elastic collisions, it is
\begin{equation}
v_{\rm min}(E_R) = \sqrt{\frac{ME_R}{2\mu^2}}.
\end{equation}
For inelastic collisions, a dark matter ground state $\chi$ may up-scatter off nuclei in the detector to an excited dark sector state $\chi^{*}$, with mass difference given by $\delta = m_{\chi^{*}} - m_{\chi}$. In this case the minimum velocity required to induce a nuclear recoil of energy $E_R$ is \cite{TuckerSmith:2001hy}
\begin{equation}
v_{\rm min}^{\rm idm}(E_R)= \frac{1}{\sqrt{2 M E_R}} \left(\frac{ME_R}{\mu} +\delta\right). \label{eq:idm_vmin}
\end{equation}

For either elastic or inelastic scattering, the observed number of events in a detector is given by the differential rate integrated over some range of recoil energies times a detector-dependent efficiency factor $\epsilon(E_R)$ and the detector exposure. 
As a result, the rate depends on experimental effects (exposure, thresholds, target mass, {\it etc}), particle physics parameters (dark matter mass and scattering cross section), and astrophysical parameters (local dark matter density and velocities). 
The particle physics information is encapsulated in the dark matter-nucleus scattering cross-section,  $\sigma_N$, in Eq.~\eqref{eq:dRdER}. To make contact with theoretical models, and to be able to compare experimental results between target nuclei, it is more convenient to write this in terms of the scattering cross section between dark matter and individual nucleons. 

In this paper, we consider two options for the dark matter-nucleon scattering: spin-independent and spin-dependent interactions. 
Though momentum-dependence can be introduced into the cross section by an appropriate choice of the particle physics interaction \cite{Feldstein:2009tr,Chang:2009yt,Feldstein:2009np,An:2010kc,Fitzpatrick:2012ix}, in this paper we assume the cross-section is independent of $q^2$, with the momentum dependence only in the form factor.

For spin-independent couplings, $\sigma_N$ can be expressed in terms of the dark matter-proton and -neutron couplings $f_p$ and $f_n$ as
\begin{equation}
\left(\sigma_N\right)_{\rm SI} = \frac{\mu^2}{\mu_p^2} \frac{\left[(A-Z)f_n + Zf_p \right]^2}{f_p^2} \sigma_p^{\rm SI} \label{eq:sigmaN_SI}
\end{equation}
where $Z$ is the atomic number, $A$ the atomic mass of the nucleus, $\mu_p$ the reduced proton-dark matter mass, and $\sigma_p^{\rm SI}$ the dark matter-proton scattering cross section.
Dark matter direct detection limits are canonically presented in terms of $\sigma_p$ assuming isospin-conservation. We will also consider isospin-violating interactions, as such interactions can change the relative signal rate between experiments with different target nuclei (with different ratios of $Z$ and $A$). For isospin-violating couplings, we parametrize the violation with the ratio
\begin{equation}
f \equiv \frac{f_n}{f_p}. \label{eq:isospin_def}
\end{equation}
and report the overall cross section in terms of the equivalent cross section $\sigma_p^{\rm SI}$. 

\begin{figure}[t]
\includegraphics[width=0.45\columnwidth]{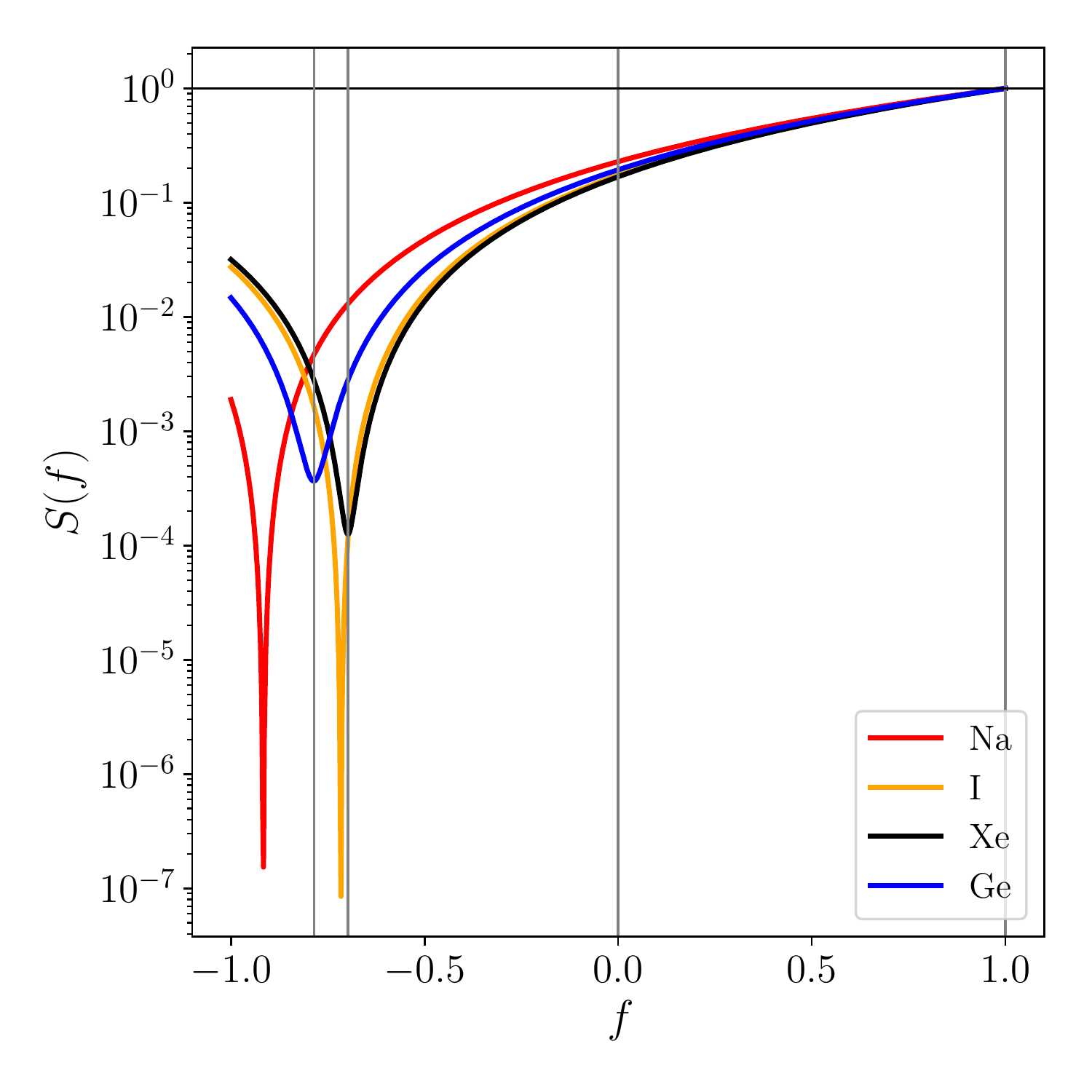}
\hspace*{0.3cm}
\includegraphics[width=0.45\columnwidth]{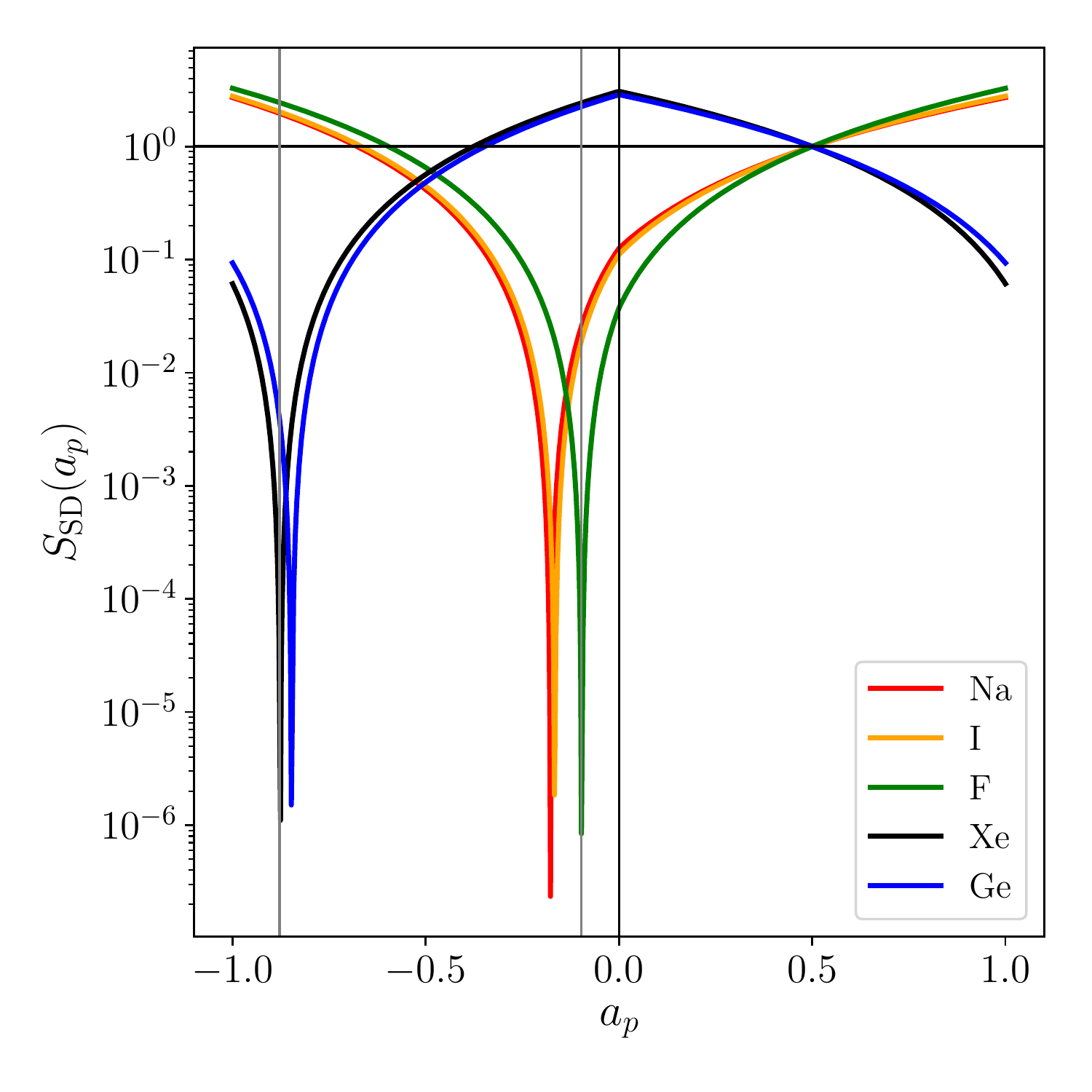}
\caption{Left: Average suppression factor $S(f)$, Eq.~\eqref{eq:SIsuppression}, for spin-independent couplings which would apply to experiments using sodium, iodine, xenon, or germanium targets, assuming natural abundances of isotopes. Vertical grey lines at $f\sim -0.700$ and $-0.785$ denote the minimum sensitivity of xenon and germanium detectors, respectively. Right: Average suppression factor $S$ for spin-dependent couplings which apply to sodium, iodine, xenon, germanium, or fluorine targets, and with the constraint $|a_p|+|a_n| =1$. Vertical grey lines as $a_p \sim -0.098$ and $-0.88$ correspond to minimum sensitivity for fluorine and xenon, respectively, see Eqs.~\eqref{eq:SDmixing} and \eqref{eq:SDsuppression}.  \label{fig:isospin_violation} }
\end{figure}

For dark matter scattering off a target that is composed of multiple isotopes (of one or more elements), the scattering rate measured would be the sum over the rates from Eq.~\eqref{eq:dRdER} for each isotope, weighted by the isotope abundance. In principle, this means that certain direct detection target materials could have greatly suppressed rates, if the isospin-violating coupling $f$ is tuned such that
\begin{equation}
f \approx -\frac{Z}{A-Z}
\end{equation}
for the target isotope \cite{Feng:2013fyw}. Even ignoring the coincidence this would imply, for many of the experiments setting the strongest current constraints the scattering cross section cannot be tuned to zero, because the natural abundance of the target elements consists of multiple isotopes. For such targets, which include xenon- and germanium-based detectors, a single value of $f$ cannot cancel away interactions with all of these nuclei. In Figure~\ref{fig:isospin_violation}, we show the abundance-averaged suppression factor -- defined as the reduction in scattering cross section relative to the isospin-conserving case -- as a function of the isospin violating parameter $f$:
\begin{equation}
S(f) = \sum_{ i} P_i \left[\frac{(A_i-Z_i)f-Z_i}{A_i}\right]^2, \label{eq:SIsuppression}
\end{equation}
where the sum runs over isotopes $i$ with abundance $P_i$, atomic numbers $Z_i$, and mass numbers $A_i$. This factor $S(f)$ gives an estimate in the reduction in sensitivity for a given experiment.

For spin-dependent interactions, the equivalent of Eq.~\eqref{eq:sigmaN_SI} can be written as
\begin{equation}
\left(\sigma_N\right)_{\rm SD} = \mu^2 \left[ a_p \langle S_p \rangle + a_n \langle S_n \rangle \right]^2 \frac{J+1}{J} = \frac{\mu^2}{\mu_p^2} \left[ \langle S_p \rangle + \frac{a_n}{a_p} \langle S_n \rangle \right]^2 \left(\frac{4}{3} \frac{J+1}{J}\right) \sigma_p^{\rm SD} \label{eq:sigmaN_SD}
\end{equation}
where $a_p$ and $a_n$ are the proton- and neutron-couplings (the spin-dependent equivalents of $f_p$ and $f_n$). The total nuclear spin is $J$, and the spin-expectation values for the proton and neutrons are $\langle S_p\rangle$ and $\langle S_n\rangle$. The spin parameters for nuclei relevant to direct detection are listed in Table~\ref{tab:nuc_params}. As nuclear spins are typically small, and Eq.~\eqref{eq:sigmaN_SD} lacks the equivalent of the $A^2$ enhancement of Eq.~\eqref{eq:sigmaN_SI}, the experimental reach on spin-dependent scattering (expressed in terms of $\sigma_p^{\rm SD}$) are much weaker than the spin-independent ones. 

Somewhat confusingly, in the direct detection literature, the ``proton'' and ``neutron'' couplings $a_p$ and $a_n$ are defined in terms of the tree-level Lagrangian couplings. Pion exchange will mix these couplings, introducing a coupling to neutrons even in the limit $a_n \rightarrow 0$, or a coupling to protons when $a_p$ is zero. We follow the calculations of Ref.~\cite{Klos:2013rwa} for the size of this mixing effect:
\begin{equation}
a_p \to a_p +\frac{1}{2}(a_p-a_n)\delta a,\quad a_n \to a_n -\frac{1}{2}(a_p-a_n)\delta a, \label{eq:SDmixing}
\end{equation}
where the momentum-dependent mixing parameter $\delta a$ is generically $\sim -20\%$. As in the spin-independent case, isospin-violation for spin-dependent interactions corresponds to a ratio $a_n/a_p \neq 1$. In the right-hand panel of Figure~\ref{fig:isospin_violation}, we show the equivalent of Eq.~\eqref{eq:SIsuppression} for spin-dependent scattering, defined as the suppression (or enhancement) of the cross section relative to the isospin-conserving assumption of $a_p = a_n$:
\begin{equation}
S_{\rm SD}(a_p) = \frac{\sum_{ i} P_i\left[ a_p\langle S_p \rangle_i + (1-|a_p|) \langle S_n \rangle_i \right]^2}{\sum_{ i} P_i\left[  \tfrac{1}{2} \langle S_p \rangle_i + \tfrac{1}{2} \langle S_n \rangle_i \right]^2}, \label{eq:SDsuppression}
\end{equation}
where we have imposed the constraint $|a_p|+|a_n|=1$ (this normalization is chosen as several nuclei relevant for spin-dependent scattering have either $\langle S_p \rangle \gg \langle S_n \rangle$ or $\langle S_p \rangle \ll \langle S_n \rangle$, making $a_n/a_p \gg 1$ and $a_n/a_p \ll 1$ both potentially interesting). Following  Ref.~\cite{Klos:2013rwa}, at zero momentum exchange, the mixing induced by pion exchange in Eq.~\eqref{eq:SDmixing} is  taken to be $\delta a \approx - 0.2$.

\begin{table}
\centering
 \begin{tabular}{| c || c | c | c | c |}
\hline
 Isotope & Abundance & $J$ & $\langle S_p \rangle$ & $\langle S_n \rangle$ \\ \hline \hline
 $^{19}$F & 1.0 & $1/2$ & $0.458$ & $-0.059$ \\ \hline
 $^{23}$Na & 1.0 & $3/2$ & $0.224$ & $0.024$  \\ \hline
  $^{29}$Si & 1.0 & 1/2 & $0.016$ & $0.156$ \\ \hline
 $^{73}$Ge & 0.0776 & 9/2 & 0.031 & 0.439  \\ \hline
 $^{127}$I & 1.0 & $5/2$ & $0.342$ & $0.031$   \\ \hline
 $^{129}$Xe & 0.264 & $1/2$ & $0.010$ & $0.329$  \\ \hline
 $^{131}$Xe & 0.212 & $3/2$ & $-0.009$ & $-0.272$  \\ \hline
  \end{tabular}
  \caption{Nuclear spins and nucleon expectation values for the different nuclei used in the experiments considered in this work. We use the proton and neutron expectation values compiled from nuclear shell models in Ref.~\cite{Klos:2013rwa} 
  \label{tab:nuc_params}}
\end{table}

Importantly, in Eq.~\eqref{eq:dRdER} the dependence of the rate on the dark matter velocity distribution can be factorized from the particle physics and experimental effects up to the dependence on $v_{\rm min}(E_R)$. Different experiments can then be compared in an ``astrophysics-independent'' manner \cite{Anderson:2015xaa, Fox:2014kua, Ibarra:2017mzt, Gelmini:2017aqe, Bozorgnia:2014gsa, DelNobile:2013cva, HerreroGarcia:2012fu, HerreroGarcia:2011aa, Gondolo:2012rs, Frandsen:2011gi} by comparing the rates observed as functions of $v_{\rm min}$. A particular dark matter direct detection experiment is sensitive to those dark matter particles moving fast enough (in the lab-frame) to scatter with a nucleus in the target, imparting a recoil energy above the detector threshold. The rate can then be expressed as a convolution between the experimental response and a piece depending on the dark matter velocity distribution \cite{Baum:2018ekm, Necib:2018iwb}, in particular the average of the inverse velocity, integrating over all velocities above an experiment-dependent $v_{\rm min}$:
\begin{equation}
\eta(v_{\rm min}, t) = \int_{v_{\rm min}}^\infty v^2 dv \int d\Omega \frac{\tilde{f}(\vec{v}, t)}{v}.
\end{equation}
The velocity distribution in the lab-frame $\tilde{f}$ must be related to the distribution in the Galactic rest frame (to which the lab is moving with a relative velocity $\vec{v}_{\rm lab}$) via $\tilde{f}(\vec{v}, t) = f(\vec{v}+\vec{v}_{\rm lab}(t))$.
In order to compare the \dama~experiment with those that only measure the time-integrated rate, a standard dark matter velocity distribution is usually adopted.
In the next section, we review this standard choice of distribution, as well as the discoveries about the local velocity distribution made possible by the \gaia~mission.


\section{Local Distribution of Dark Matter \label{sec:distribution}}
\subsection{Dark Matter in the Halo}

\begin{figure}[t]
\includegraphics[width=0.45\columnwidth]{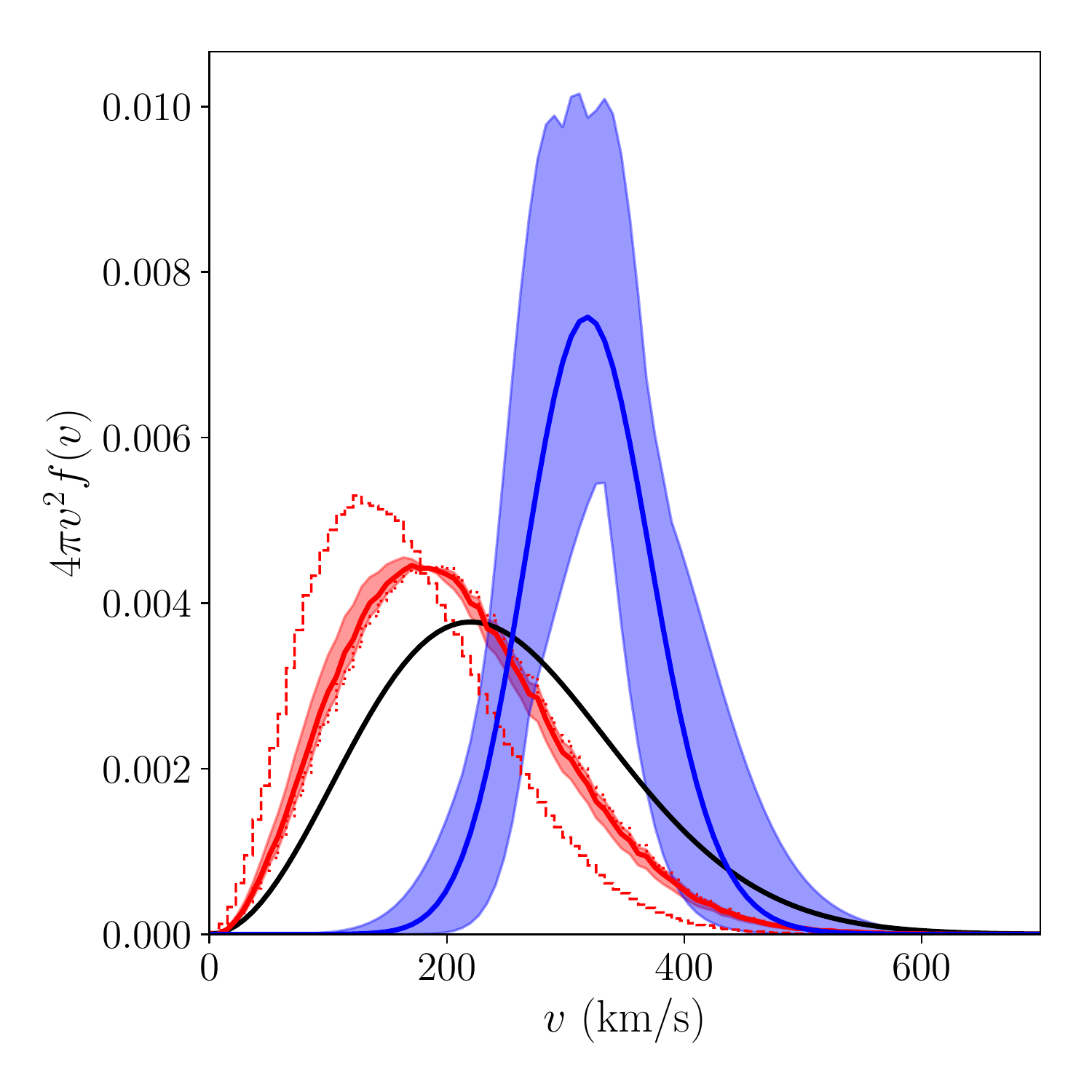}
\hspace*{0.3cm}
\includegraphics[width=0.45\columnwidth]{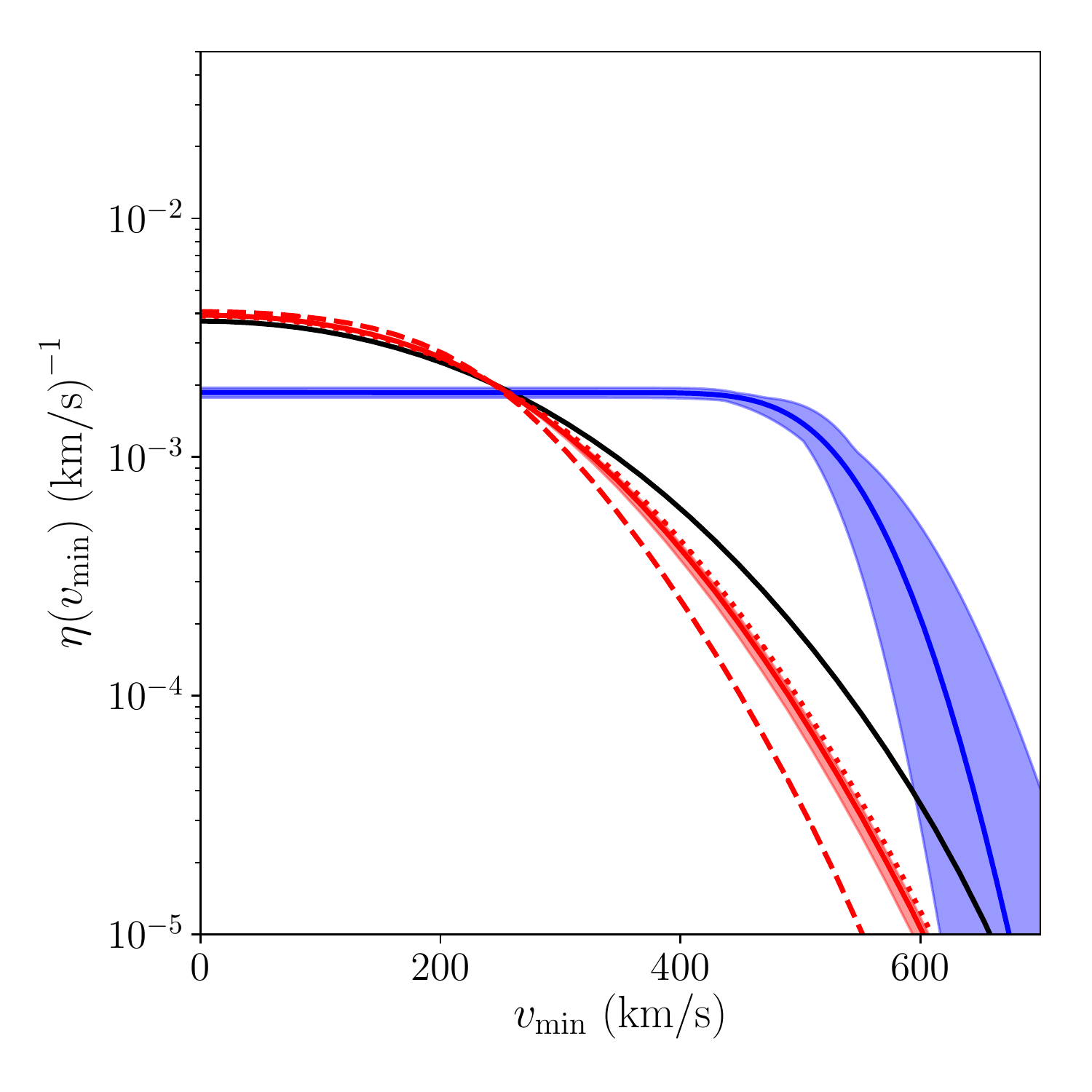}
\caption{Normalized dark matter velocity distributions in the Galactic rest frame (left) and $\eta(v_{\rm min})$ in the Earth frame averaged over one year (right) for the Standard Halo Model (black line), the model derived by Ref.~\cite{Necib:2018igl} from \gaia~data for the halo (red dotted), substructure (red dashed) and sum (red solid lines, with shaded region indicating uncertainties). The blue line is the normalized velocity distribution for the S1 stream, with shading indicating uncertainties. Recall that, though we normalize all distributions to unity in these plots, the stream will not contain 100\% of the local dark matter density. \label{fig:DMdistributions} }
\end{figure}

The simplest assumption of the local motion of dark matter is that the local dark matter velocity in the Galactic rest frame is an isotropic Maxwell-Boltzmann distribution -- the Standard Halo Model (SHM):
\begin{equation}
f(\vec{v}) = \frac{1}{N} \frac{1}{(\pi v_0)^{3/2}} e^{-v^2/v_0^2}\Theta(v_{\rm esc}-|v|). \label{eq:SHMf}
\end{equation}
where the normalization factor is
\begin{equation}
N = \mbox{erf}\left[\frac{v_{\rm esc}}{v_0} \right] - \frac{2v_{\rm esc}}{\sqrt{\pi}v_0} e^{-v_{\rm esc}^2/v_0^2}.
\end{equation}
We adopt the SHM parameters $v_0 = 220$~km/s for the dark matter mean speed, and a Galactic escape velocity of $v_{\rm esc} = 544$~km/s \cite{Smith:2006ym}. The SHM distribution $f(v)\times 4\pi v^2$ in the Galactic rest-frame is shown in the left panel of Figure~\ref{fig:DMdistributions}.

To make contact with experiments, which are sensitive to the lab-frame $\tilde{f}$ rather than the rest-frame $f$, we must also specify the relative velocity of the Earth and Sun with respect to the Galactic standard of rest. In Galactic coordinates ($x$ radial from the Galactic Center, $z$ out of the disk), the Sun's velocity is
\begin{equation}
\vec{v}_\odot = (10,233,7)~{\rm km/s}. \label{eq:sunvelocity}
\end{equation}
To this, we must add the time-dependent Earth velocity \cite{Savage:2009mk}
\begin{eqnarray}
\vec{v}_\oplus & = & (29.8~{\rm km/s})\left[ \hat{\epsilon}_1 \cos\left(\omega(t-t_1) \right) + \hat{\epsilon}_2 \sin\left(\omega(t-t_1) \right) \right], \label{eq:earthvelocity} \\
\hat{\epsilon}_1 & = & (0.993,0.117,-0.010), \\ 
\hat{\epsilon}_2 & = & (-0.067,0.493,-0.868),
\end{eqnarray}
where $\omega = 2\pi/365.25$~days, with the phase shift $t_1 = 79.62$ (relative to January 1$^{\rm st}$). The annual modulation of the total velocity of the Earth+Sun system will modify the integral over $\tilde{f}(v)$ for a fixed $v_{\rm min}$, thus causing a yearly modulation in a direct detection experiment's sensitivity and resulting number of events, as reported by \dama.

\begin{figure}[t]
\includegraphics[width=0.425\columnwidth]{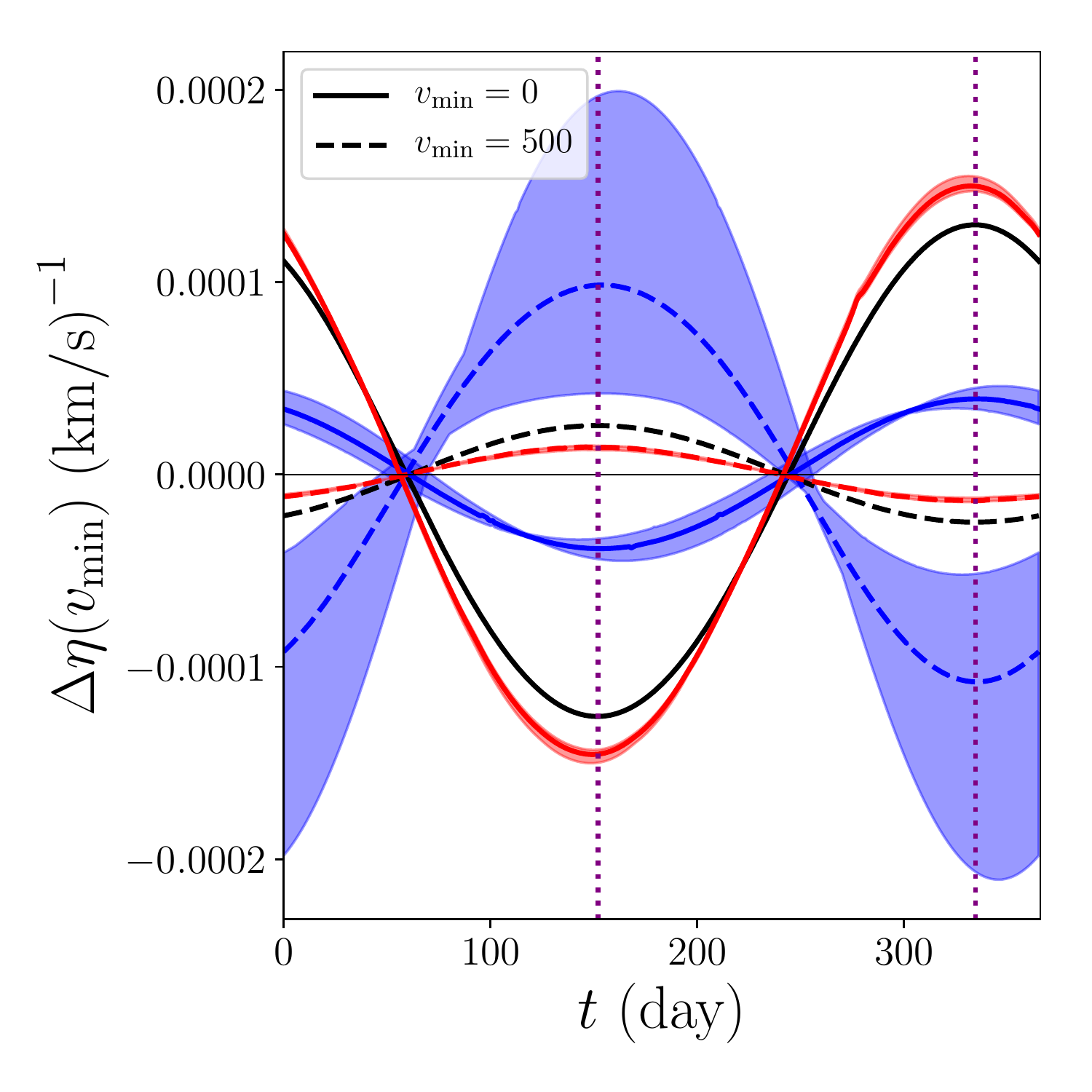}
\hspace*{0.3cm}
\includegraphics[width=0.46\columnwidth]{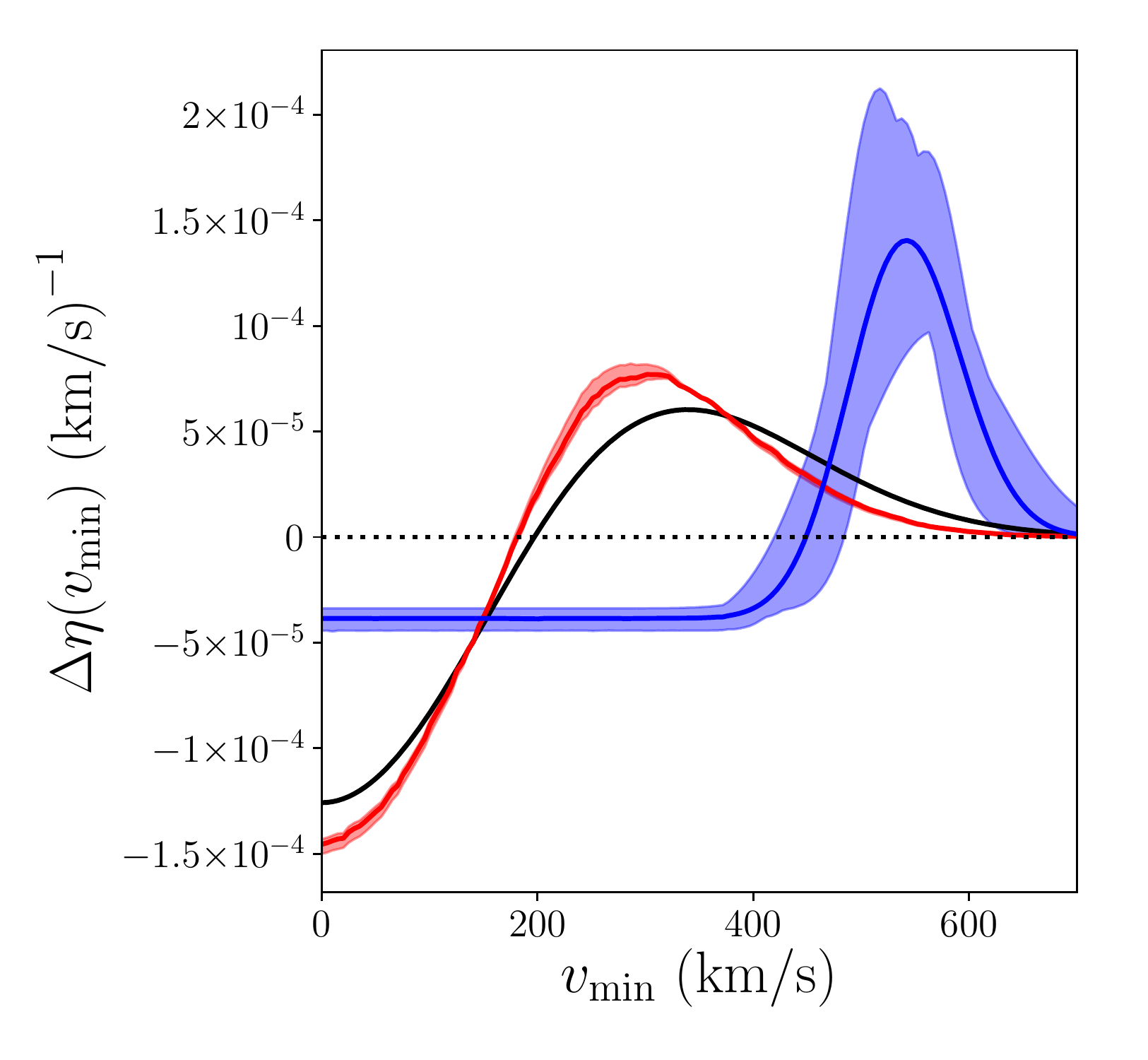}
\caption{Left: Modulation of the $\Delta \eta \equiv \eta(v_{\rm min},t) - \bar{\eta}(v_{\rm min})$ as a function of time over a year for $v_{\rm min} = 0$ and $500$~km/s. Right: the maximum $\Delta \eta(v_{\rm min})$ as a function of $v_{\rm min}$ (defined such that positive $\Delta \eta$ corresponds to peaks near $t\approx 152$~days). The SHM modulation is the black solid line, the \gaia-derived model is the red solid line with uncertainties indicated by the shaded region. The S1 stream best-fit is the  blue solid line, with uncertainties shaded. The day $t=0$ corresponds to January 1$^{\rm st}$, with the predicted peak days for the SHM (for high $v_{\rm min}$) shown with vertical dotted lines. \label{fig:eta_modulation} }
\end{figure}

For a dark matter distribution that can be modeled as an isotropic Maxwell-Boltzmann in the Galactic rest frame, such as the SHM, the $\eta$ function has an analytic form (see e.g.~Ref.~\cite{Savage:2009mk})
\begin{equation}
\eta(v_{\rm min}) = \left\{\begin{array}{lc} \frac{1}{v_{\rm lab}}, & v_{\rm esc} < v_{\rm lab},\,v_{\rm min} < |v_{\rm lab} - v_{\rm esc}| \\
\frac{1}{2N v_{\rm lab}}\left[\mbox{erf}\left( \frac{v_{\rm min}+ v_{\rm lab}}{v_0}\right) - \mbox{erf}\left( \frac{v_{\rm min}- v_{\rm lab}}{v_0} \right) - \frac{4v_{\rm lab}}{\sqrt{\pi} v_0}e^{-v_{\rm esc}^2/v_0^2}\right] , & v_{\rm esc} > v_{\rm lab},\,v_{\rm min} < |v_{\rm lab} - v_{\rm esc}| \\
\frac{1}{2N v_{\rm lab}}\left[\mbox{erf}\left( \frac{v_{\rm esc}}{v_0}\right) - \mbox{erf}\left( \frac{v_{\rm min}- v_{\rm lab}}{v_0} \right) - \frac{2}{\sqrt{\pi} v_0}(v_{\rm lab} + v_{\rm esc} - v_{\rm min})e^{-v_{\rm esc}^2/v_0^2}\right] , & |v_{\rm lab} - v_{\rm esc}| < v_{\rm min} < v_{\rm lab} + v_{\rm esc} \\
0, & v_{\rm lab} + v_{\rm esc} < v_{\rm min} \end{array} \right.
\end{equation}
The $\eta$ function of the SHM with the benchmark parameters is shown in the right panel of Figure~\ref{fig:DMdistributions}, while the modulation of $\eta$ over the year is shown in Figure~\ref{fig:eta_modulation}. Notice (in the left panel of Figure~\ref{fig:eta_modulation}) that the modulation $\Delta \eta(t) \equiv \eta(t) - \bar{\eta}$ switches sign as $v_{\rm min}$ is increased, from a minimum around June 2$^{\rm nd}$ to a maximum on that same day. 

However, it has long been known \cite{Butsky:2015pya,Kuhlen:2009vh,Kuhlen:2013tra,Ling:2009eh,Lisanti:2010qx,Sloane:2016kyi,Vogelsberger:2008qb,Bozorgnia:2016ogo,Kelso:2016qqj} that the true distribution of dark matter in the Milky Way-like halos should deviate significantly from this idealized distribution.
Dark matter direct detection experiments often take this uncertainty into account when reporting limits \cite{McCabe:2010zh,Fowlie:2018svr, Wu:2019nhd, Agnese:2017jvy}. 

With the \gaia~DR2, our expectation that dark matter should deviate from the SHM can be directly tested, and a halo distribution inferred using low-metallicity stars in the Galactic halo \cite{Necib:2018iwb,Necib:2018igl}. Thus, the astrophysical uncertainty in direct detection experiments can be significantly reduced. We adopt the distributions of Ref.~\cite{Necib:2018igl} for the model of the dark matter halo. We also note the recent work of Ref.~\cite{Wu:2019nhd}, which also derived modifications to the dark matter velocity distribution and the resulting implications on direct detection from \gaia~data; in that case the primary effect coming from the changes in the escape velocity of the halo.

The \gaia-derived distributions used in this work have two components: dark matter from a smooth halo which has been integrated into the Galaxy, and a substructure of dark matter from tidal debris \cite{Lisanti:2011as,Lisanti:2014dva,Necib:2018igl}, which is likely the remnant in velocity-space of long-ago major merger events. The total dark matter distribution in the local volume of the Milky Way is then $c_{\rm halo} f_{\rm halo} + c_{\rm sub} f_{\rm sub}$ with \cite{Necib:2018igl}
\begin{equation}
\frac{c_{\rm sub}}{c_{\rm halo}} = 0.23^{+0.43}_{-0.15}. \label{eq:substructureratio}
\end{equation}
The $f_{\rm halo}(v)$ and $f_{\rm sub}(v)$ are the velocity distributions derived from the halo and substructure respectively, which we adopt from Ref.~\cite{Necib:2018iwb} \footnote{We thank the authors of Ref.~\cite{Necib:2018iwb} for providing their velocity distribution data files.}. The \gaia-derived distributions however do not take into account the contribution from dark matter substructure or smooth accretion, both of which could be non-negligible as pointed out in Refs.~\cite{Springel:2008cc, Sawala:2014baa, Evans:2018bqy, Bozorgnia:2018pfa}.
In Figure~\ref{fig:DMdistributions} we show the SHM velocity distribution and $\eta$ functions with these data-driven distributions. The data-driven model of the velocity distribution has fewer high-velocity particles, which lead to weaker bounds for light dark matter (since as the dark matter mass is lowered, $v_{\rm min}$ correspondingly increases). Furthermore, as mentioned above, Ref.~\cite{Evans:2018bqy} derived a smooth halo distribution with an excess at higher velocities, which though was found to have little impact on current dark matter limits, did lead to marginally stronger bounds compared to the SHM in the low mass region. 
Similarly, as seen in Figure~\ref{fig:DMdistributions}, the amplitude of $\Delta \eta$ decreases for $v_{\rm min} \gtrsim 350$~km/s for this \gaia-derived halo model (though the peak day is nearly unchanged).
Given the four to six orders of magnitude between the \dama~best fit region and the \xenon~limits (assuming spin-independent isospin-conserving couplings), the \dama~region is safely ruled out for modifications of the smooth dark matter velocity distribution (assuming velocity-independent cross sections), even with these slightly weakened limits (see, e.g., Ref.~\cite{Sloane:2016kyi}). 

\subsection{The S1 Stream}

The possibility that the Earth lies in the path of multiple kinematically cold streams of dark matter has long been known. Such streams will modify the dark matter velocity distribution in ways not captured by simulations of the overall halo and can have large effects on the modulation measured in direct detection \cite{Savage:2006qr}. 
The stream S1, identified in Ref.~\cite{Myeong:2017skt}, provides a concrete example and requires us to revisit the conclusion that the null results of other experiments definitively exclude the \dama~best-fit regions. S1 is a very high velocity stream, with a velocity of $\sim 300$~km/s relative to the Galactic rest frame. These stars are coherent in velocity space, and though fairly spread out in position space, do overlap with the position of the Sun within the Milky Way. Such ``tidal debris'' is expected from simulations \cite{Lisanti:2011as,Lisanti:2014dva,Necib:2018igl}, and the stream S1 can provisionally be identified as the stellar remnants of a $\sim 2\times 10^{10}\,M_\odot$ dwarf galaxy which was absorbed by the Milky Way some 10 billion years ago. The contribution of the S1 stream to the local dark matter density is not known at this point, and in this paper we will treat it as a free parameter. Note that the S1 stream is within the Galactic disk, and so was not part of the \gaia~stellar sample that Ref.~\cite{Necib:2018igl} used to construct their model of the overall halo distribution.

Relevant for the \dama~experiment, the stream is essentially anti-parallel to the Sun's motion through the Milky Way. This is important for two reasons. First, as a result of this antiparallel motion, the local dark matter wind from the stream has a very large relative velocity, which can increase the magnitude of a modulation signal. Second, as the Earth moves around the Sun, the velocity distribution of the dark matter stream peaks very close to the date one would predict from the SHM alone.

The \dama~search (and claimed positive signal) is reliant on the yearly modulation of the dark matter $\eta$ function, due to the Earth's motion around the Sun. In order to match the observed signal, the modulation signal must peak around June $2^{\rm nd}$ ($t = 152.5$). This is close to the expected signal peak for a SHM dark matter distribution, as it is close to the day that Earth's velocity relative to Galactic rest is maximized (as noted previously, the improved dark matter distributions derived from \gaia~data have the same peak day). 


Though the density is not known, the velocities of the stars within S1 are known, and from this, a velocity distribution can be modeled. We assume in this paper that the low and intermediate-metallicity stars are good tracers of the component of the dark matter in the halo accreted at early times from luminous satellites; this conclusion has been supported by $N$-body simulation \cite{Necib:2018igl}, though simulation also demonstrates that there remains a component of dark matter in the halo which is not traced by the stars \cite{Bozorgnia:2018pfa}. Certainly one should expect streams of dark matter to exist without accompanying stars, but not necessarily the reverse: stars from tidally disrupted dwarf galaxies should be accompanied by dark matter. 

The stars in S1 are counter-rotating relative to the Sun, with mean velocity $\vec{v}_{S1} = (8.6,-286.7,-67.9)$~km/s in Galactic cylindrical coordinates. Assuming that the velocity distribution of the stars in the Galactic rest frame is drawn from a three-dimensional Maxwell-Boltzmann distribution with diagonal velocity dispersion matrix $v_0$:
\begin{equation}
f(\vec{v}) = \frac{1}{\sqrt{\pi^3 {\rm det} (v_0)^2}} {\rm exp}\left[-(\vec{v} - \vec{v}_{S1})\cdot \frac{1}{(v_0)^2}\cdot (\vec{v} - \vec{v}_{S1})\right],
\end{equation}
then, following the work of Ref.~\cite{OHare:2018trr}\footnote{We thank the authors of Ref.~\cite{OHare:2018trr} for kindly providing their kinematic data, allowing us to refit and extract errors.}, we find
\begin{eqnarray}
\vec{v}_{S1} & = & (8.6\pm37.2,-286.7\pm18.5,-67.9\pm22.1)~{\rm km/s}, \label{eq:S1parameters} \\
v_0 & = & {\rm diag}\left[163.1\pm 37.1,70.6\pm 18.4,84.5\pm 22.2\right]~{\rm km/s}. \nonumber
\end{eqnarray}
The $f(v)$ and $\eta$ distribution for the S1 stream are shown in Figure~\ref{fig:DMdistributions}, and the yearly modulation of $\eta$ is shown in Figure~\ref{fig:eta_modulation}. Notice that for large $v_{\rm min} \gtrsim 450$~km/s, the magnitude of $\Delta \eta$ can be nearly an order of magnitude higher than in either the SHM or the data-driven model. 
While $\eta$ itself also increases, for very high $v_{\rm min}\sim 500-600$~km/s, the relative increase in $\Delta \eta$ as compared to $\eta$ means that it is possible for experiments which measure modulation to get very large boosts in sensitivity while those that measure only average rate would be comparatively unchanged. Furthermore, even if the stream contributes only ${\cal O}(10\%)$ to the local dark matter density, the sharp feature in the recoil spectrum it could induce could be visible in a direct detection experiment, and change the particle physics parameters of a best-fit point.  It is these relative changes in signal sensitivity due to the S1 stream that we are interested in here.
Also shown in Figures ~\ref{fig:DMdistributions} and \ref{fig:eta_modulation} are the uncertainties; the red shaded regions correspond to uncertainties over the substructure/halo parameter fit given by Eq.~\ref{eq:substructureratio}, while the blue shaded region represents the uncertainties in the stream parameters, given by Eq.~\ref{eq:S1parameters}.

\section{Experimental Limits \label{sec:limits}}
We are now in the position to reassess the current experimental limits in light of our new understanding of the local dark matter velocity distribution. 

We are interested in determining the exclusion reach and signal regions of the existing direct detection experiments in light of the \gaia-derived halo mode and in the presence of the high-velocity S1 stream. In particular, does dark matter moving in this stream allow the \dama~best-fit region to evade the other experiment's null constraints?
The presence of the S1 stream induces features in the recoil spectrum that would not be expected in a smooth halo model, and so may be of experimental interest. As astrometric surveys of the local neighborhood are still in their early days, it may be expected that more streams and tidal debris will be identified in the future, in which case this study of the S1 stream can provide an example of what effects on the results of direct detection experiments can be expected. As we do not know the fraction of the local dark matter density which results from the S1 stream 
we will treat this as a free parameter; the density of the S1 stream may be better constrained with future astrometric measurements and comparison with simulations. 

We consider the current strongest constraints on dark matter spin-independent and spin-dependent scattering, for dark matter masses between $1$ and $10^4$~GeV. Under the assumption of the SHM, for spin-independent searches, the strongest constraints are set by \xenon~\cite{Aprile:2018dbl} for most of the dark matter mass range, surpassed by the germanium-based \cdms~detector \cite{Agnese:2017jvy} in the low mass region. The \cosine~\cite{Adhikari2018} experiment does not set the world-leading limits for any mass range, however, we include it in this analysis as it is composed of the same target material as \dama: sodium-iodide crystals. This is important as we consider isospin-violating couplings, which can change the relative strength of the scattering of dark matter against the different target materials. The strongest limits on spin-dependent scattering are set by the \pico~experiment \cite{Amole:2017dex} for scattering against the proton, while the \xenon~limits can be reinterpreted in terms of spin-dependent scattering to give the strongest limits for spin-dependent scattering against the neutron \cite{Aprile:2019dbj}. \\

In order to extract limits (or for \dama, best-fit signal regions) for velocity profiles other than the SHM, we must recast each of these experimental results. Our procedure for each experiment is explained in  Appendix~\ref{app:recast}. Our methods do not recover the exact experimental results: for the low mass region our exclusion regions are somewhat weaker than the official limits.
As this is the mass range that will be of interest in comparing with \dama, our results are in that sense conservative. We show the 90\% CL upper limits from each experiment under various assumptions of scattering interactions and velocity distribution.

For the \dama~fits, we are interested in both the spectrum of the recoil energies and the yearly modulation. For the former, data in narrow energy bins have been made available by \dama, but these assume a yearly modulation peaking on the day predicted by the SHM. Data binned in time-series are only available for a few overlapping energy bins. We therefore provide two sets of fits to \dama: ``Amplitude'' fits to the binned recoil spectrum (where the binning in time assumes a yearly modulation), and ``modulation'' fits to the time series data, binning recoil events between $1$ and $6$ keV$_{\rm ee}$. As will be seen, the modulation fits demonstrate that the S1 stream is consistent with the same peak date as the smooth halo, and so the amplitude fits can be used for more fine-grained distinction. For both modulation and amplitude fits, we fit the signal hypothesis by minimizing a $\chi^2$ fit to the available data. Unless otherwise noted, we will show the region of parameter space that is within $2\sigma$ of the minimum $\chi^2$, taking into account the appropriate number of degrees of freedom.

\subsection{Spin-Independent Elastic Scattering}

\begin{figure}[t]
\includegraphics[width=0.45\columnwidth]{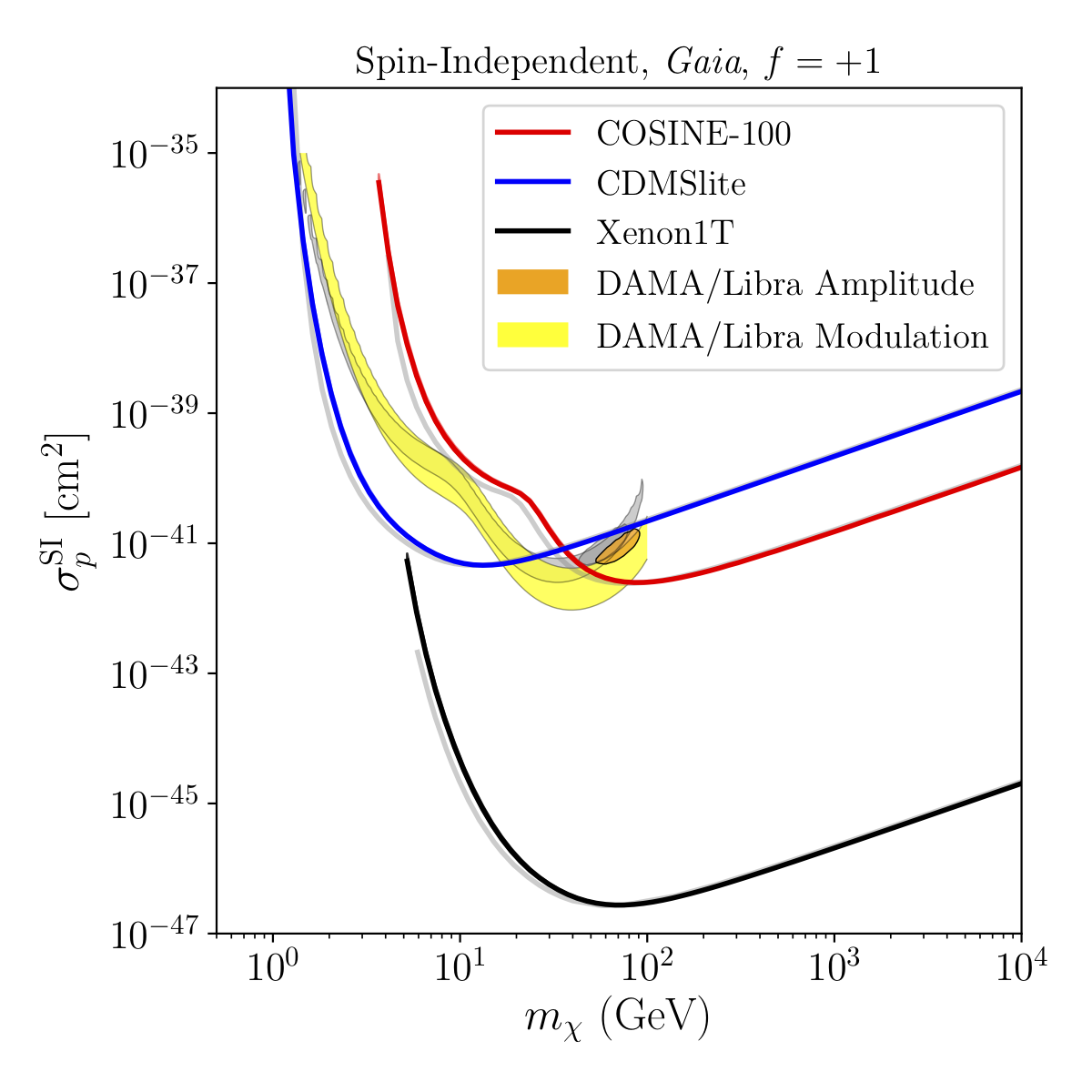}
\hspace*{0.3cm}
\includegraphics[width=0.45\columnwidth]{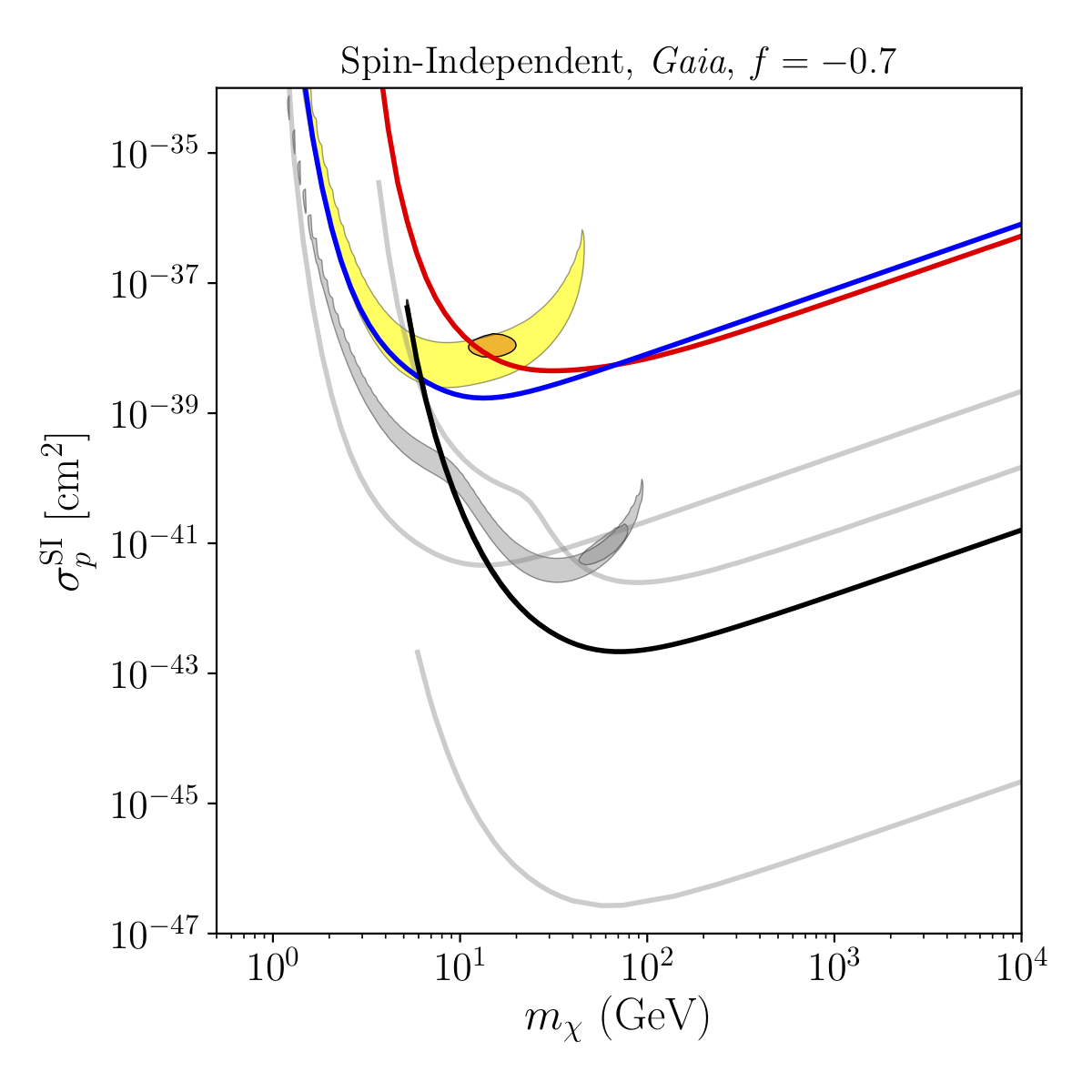}
\caption{90\%CL upper limits from \xenon~(black), \cdms~(blue) and \cosine~(red) on the spin-independent dark matter-proton scattering cross section $\sigma_p^{\rm SI}$ assuming the \gaia~halo model for isospin-conserving interactions $f=+1$ (left) and the isospin-violating interaction $f=-0.7$ (right). Best-fit $2\sigma$ regions to the \dama~data are shown in orange for fits to the reported modulation amplitudes, and in yellow for fits to the reported annual modulation rates. Equivalent isospin-conserving limits assuming the SHM are shown in shaded grey, for comparison purposes. \label{fig:nostream_isospinconserving}}
\end{figure}

Combining the limits and best-fit regions for all the experiments, we show in the left panel of Figure~\ref{fig:nostream_isospinconserving} the isospin-conserving limits assuming the data-driven dark matter velocity distribution derived from \gaia~data in Ref.~\cite{Necib:2018igl}. For each limit curve, the substructure ratio $c_{\rm sub}/c_{\rm halo}$ is varied over the $1\sigma$ range of Eq.~\eqref{eq:substructureratio}, while for the fits to \dama, this is treated as a free parameter and is allowed to float with the interaction cross-section and dark matter mass. The limits from the SHM are also displayed. 
As can be seen, \xenon~sets the strongest limits over the majority of the mass range, with \cdms~taking over below $\sim 5$~GeV. \dama~is decisively excluded by \xenon, with \cosine~excluding the claimed signal for the entire $2\sigma$ region as well. For the fit to the binned amplitude data, the best-fit \dama~point with $f = +1$ has $\chi^2/{\rm d.o.f.}=3.5$ (treating $m_\chi$, $\sigma_{p}^{\rm SI}$, and $c_{\rm sub}/c_{\rm halo}$ as free parameters), and the $2\sigma$ regions are relative to this minimum. 

\begin{figure}[t]
\includegraphics[width=0.45\columnwidth]{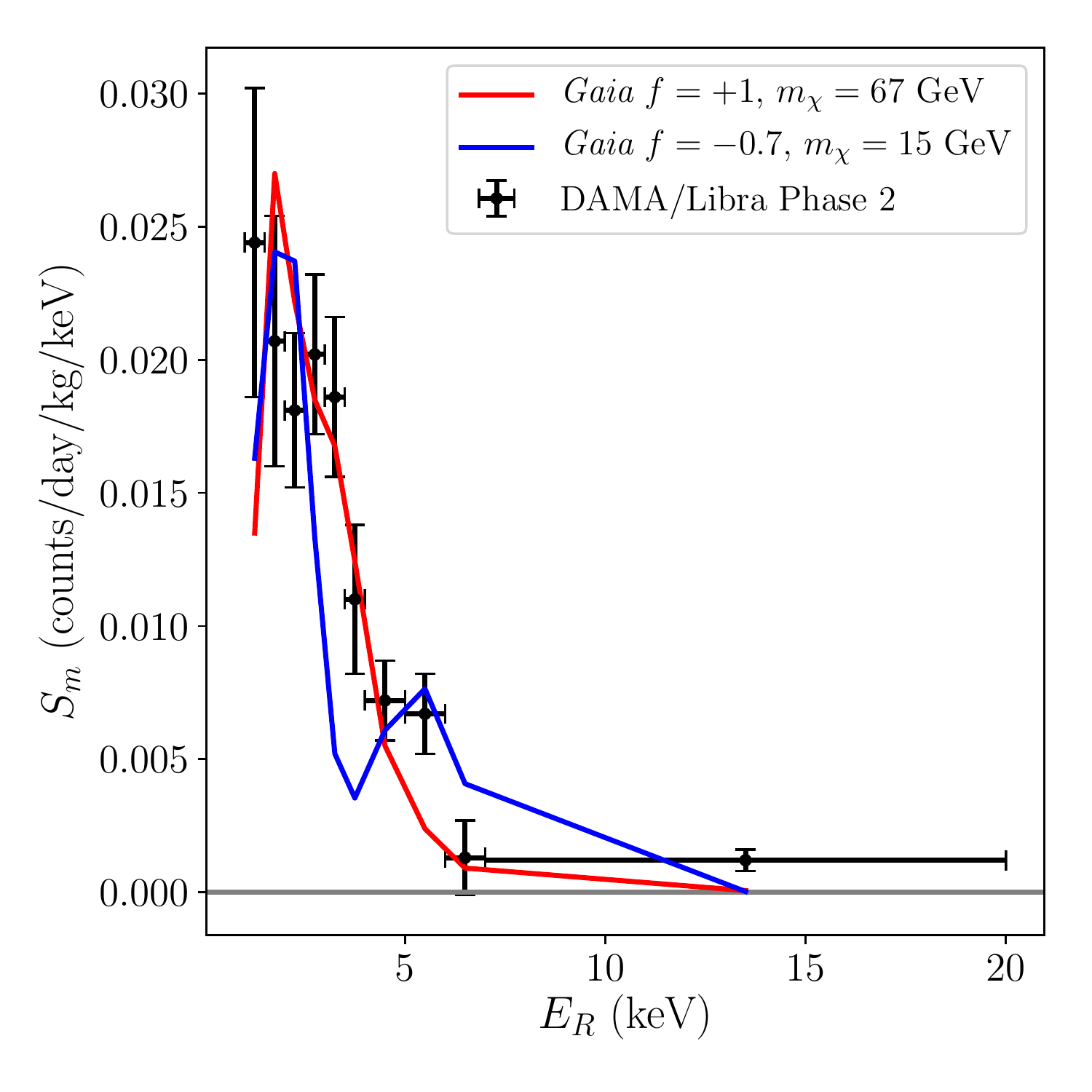}
\includegraphics[width=0.45\columnwidth]{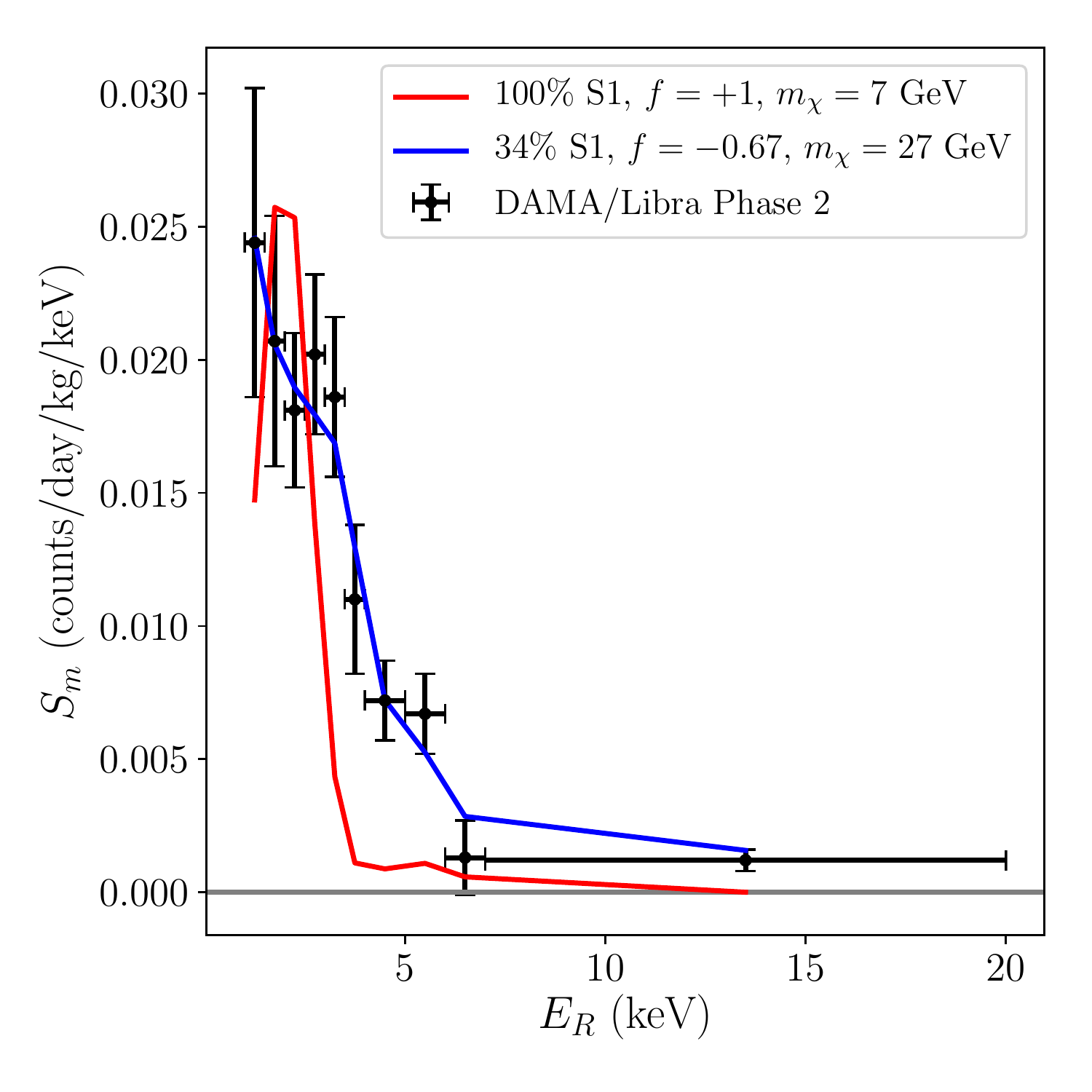}
\caption{Left: \dama~modulation amplitude Phase 2 data binned in recoil energy, from Ref.~\cite{Bernabei:2018yyw}, assuming a yearly sinusoidal modulation. Our best-fit spectrum to the Phase-2 data assuming the \gaia~halo model and $f = +1$ with best-fit dark matter mass $m_\chi = 67$~GeV is shown in red, corresponding to $\chi^2/{\rm d.o.f.} = 3.5$. Varying $f$, a lower mass fit can be found, and is shown in blue for comparison, with $m_\chi = 15$~GeV, $f=-0.7$, and $\chi^2/{\rm d.o.f.} = 7.4$. Right: Fits including the S1 stream. A best fit point assuming the stream is 100\% of the local density and $f=+1$ with $m_\chi = 7$~GeV $\chi^2/{\rm d.o.f} = 8$ is in red, and a representative good-fit parameter point for 34\% local stream density is shown in blue, corresponding to $m_\chi = 27$~GeV, $f=-0.67$ and $\chi^2/{\rm d.o.f.} = 1.2$.
\label{fig:DAMA_bestfit} }
\end{figure}

\begin{figure}[t]
\includegraphics[width=0.495\columnwidth]{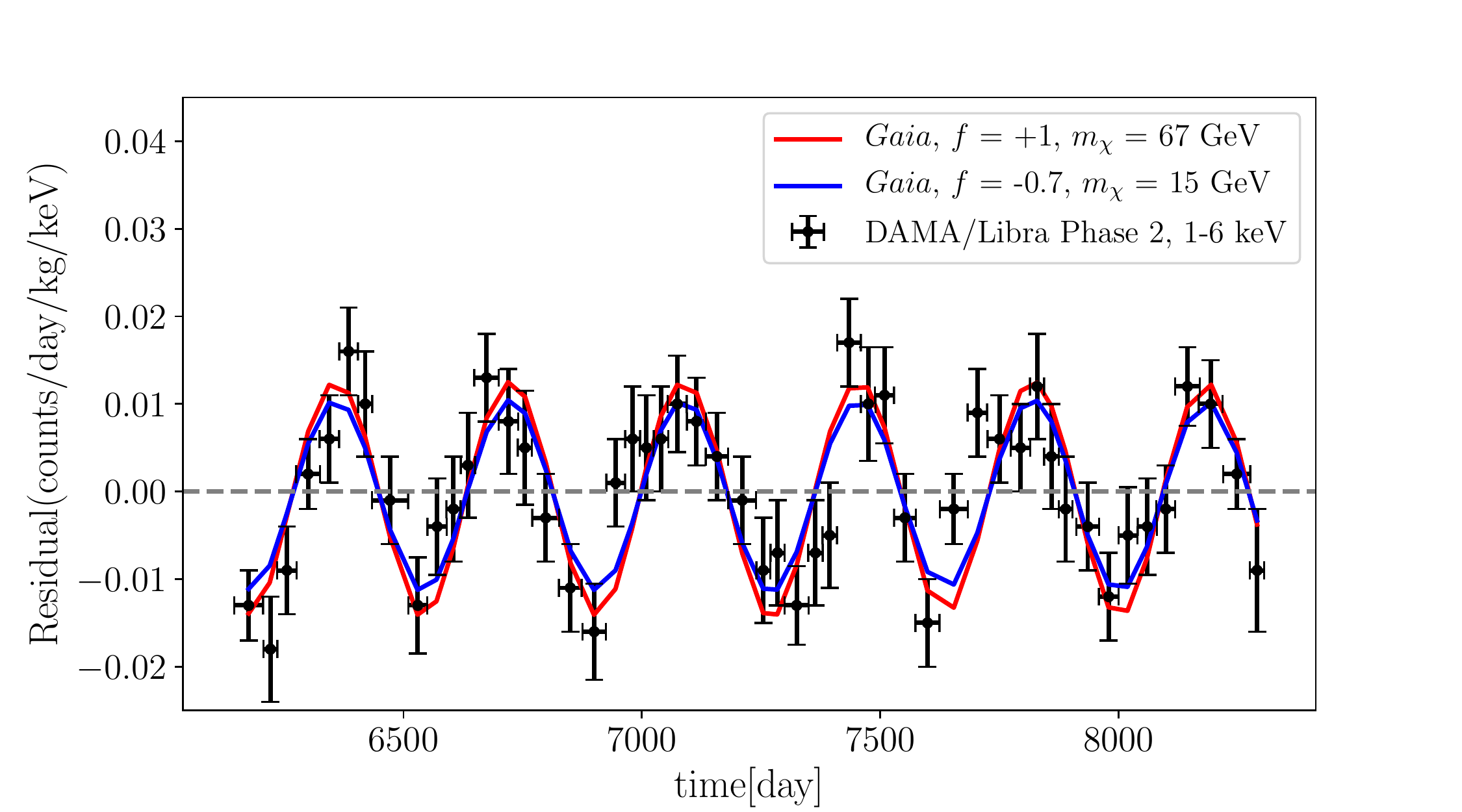}
\includegraphics[width=0.495\columnwidth]{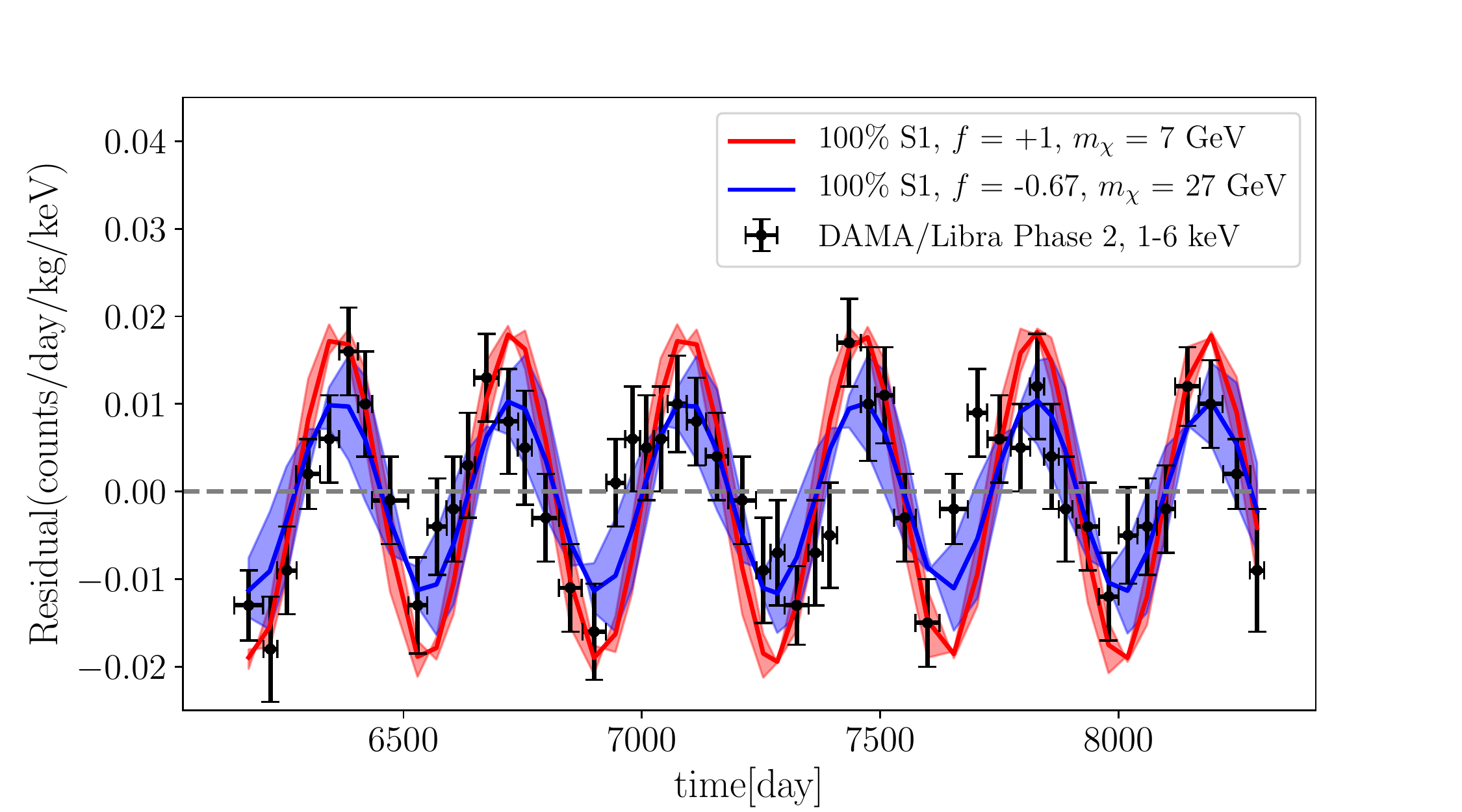}
\caption{\dama~annual modulation Phase 2 data for the $1 - 6$ keV$_{\rm ee}$ energy range, from Ref.~\cite{Bernabei:2018yyw}, along with the predicted modulation rate using the fit parameters from Figure~\ref{fig:DAMA_bestfit} above. Left: We plot the modulation rates assuming the \gaia~derived halo distribution for $f = +1$ with best fit $m_{\chi} = 67$ GeV in red and $f = -0.7$ with best fit $m_{\chi} = 15$ GeV shown in blue. Right: modulation rates including the S1 stream, shading indicates uncertainties. The red line represents $f = +1$ with best fit $m_{\chi} = 7$ GeV while $f = -0.67$ and bestfit $m_{\chi} = 27$ GeV is shown in blue. In both lines we assume the extreme case of 100\% stream density.
\label{fig:DAMA_bestfit_mod} }
\end{figure}

In the right panel of Figure~\ref{fig:nostream_isospinconserving}, we show the limits assuming the isospin violating parameter $f = -0.7$ (the value that minimizes the sensitivity of xenon-based detectors, illustrated in Figure~\ref{fig:isospin_violation}). As can be seen, changing this parameter also has the effect of suppressing scattering on iodine and enhancing sodium scattering, moving the \dama~best fit region of the amplitude data from $m_\chi \sim 60$~GeV to $m_\chi \sim 15$~GeV. The good-fit regions to the annual modulation data are consistent with those of the amplitude data, but much less constraining.
This is not surprising, given that the modulation data is binned much less finely in recoil energy, which is where most of the discriminating power in mass comes from. Given this disparity, we will mostly refer to the good-fit regions for the amplitude data, using the modulation fits to demonstrate the important point that the non-SHM velocity distributions continue to have a peak day which can generally fit the \dama~observations.

Plotting the \dama~$2\sigma$ regions relative to the minimum $\chi^2$ assuming a given value of $f$ obscures how good the overall fit is for the assumed parameter of each plot. 
To investigate this, we show in Figure~\ref{fig:DAMA_bestfit} both the best overall fit with $f= +1$ as well as the fit with $m_\chi =15$~GeV and $f=-0.7$. This point is a good fit to the annual modulation data, but a significantly worse fit to the recoil energy spectrum, with $\chi^2/{\rm d.o.f.} = 7.4$. The poor fit of the low mass \dama~region to the Phase-2 spectrum assuming the SHM has been noted previously \cite{Baum:2018ekm}. The resulting predictions for the annual modulation of the event rate, compared to the published Phase-2 data, is shown in Figure~\ref{fig:DAMA_bestfit_mod}, which demonstrates that the \gaia~halo models have the appropriate phase shift to match the observations.

\begin{figure}[t]
\includegraphics[width=0.325\columnwidth]{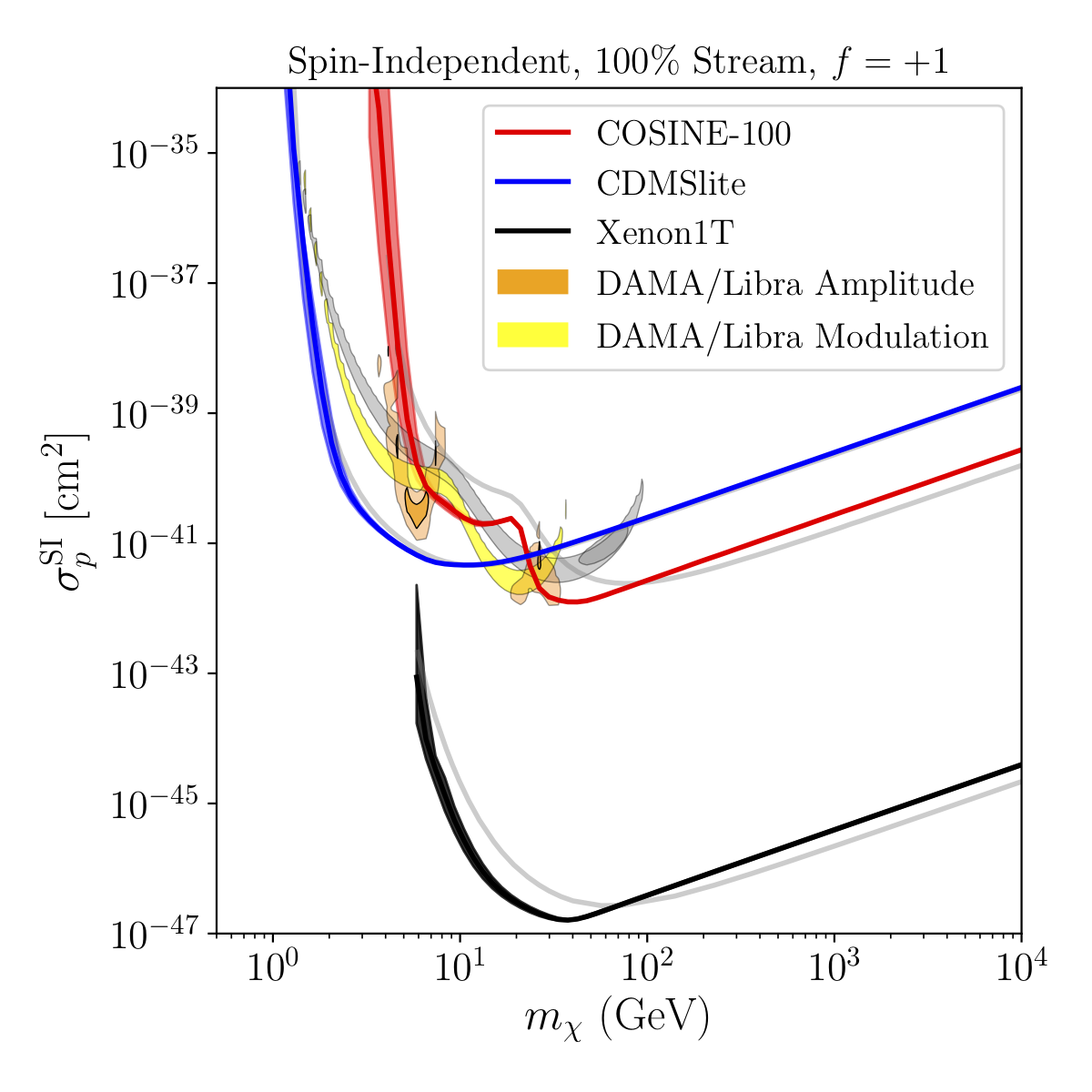}
\includegraphics[width=0.325\columnwidth]{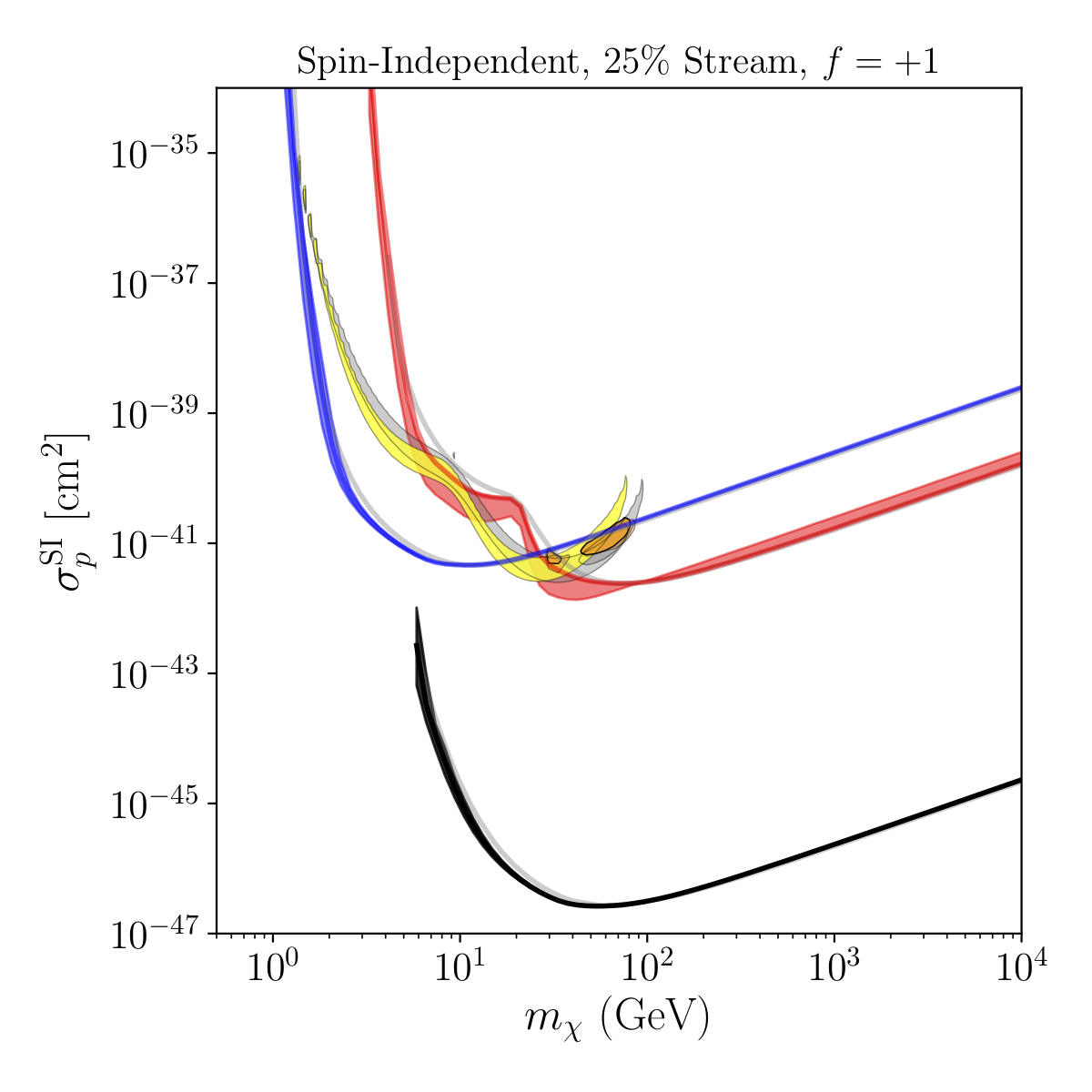}
\includegraphics[width=0.325\columnwidth]{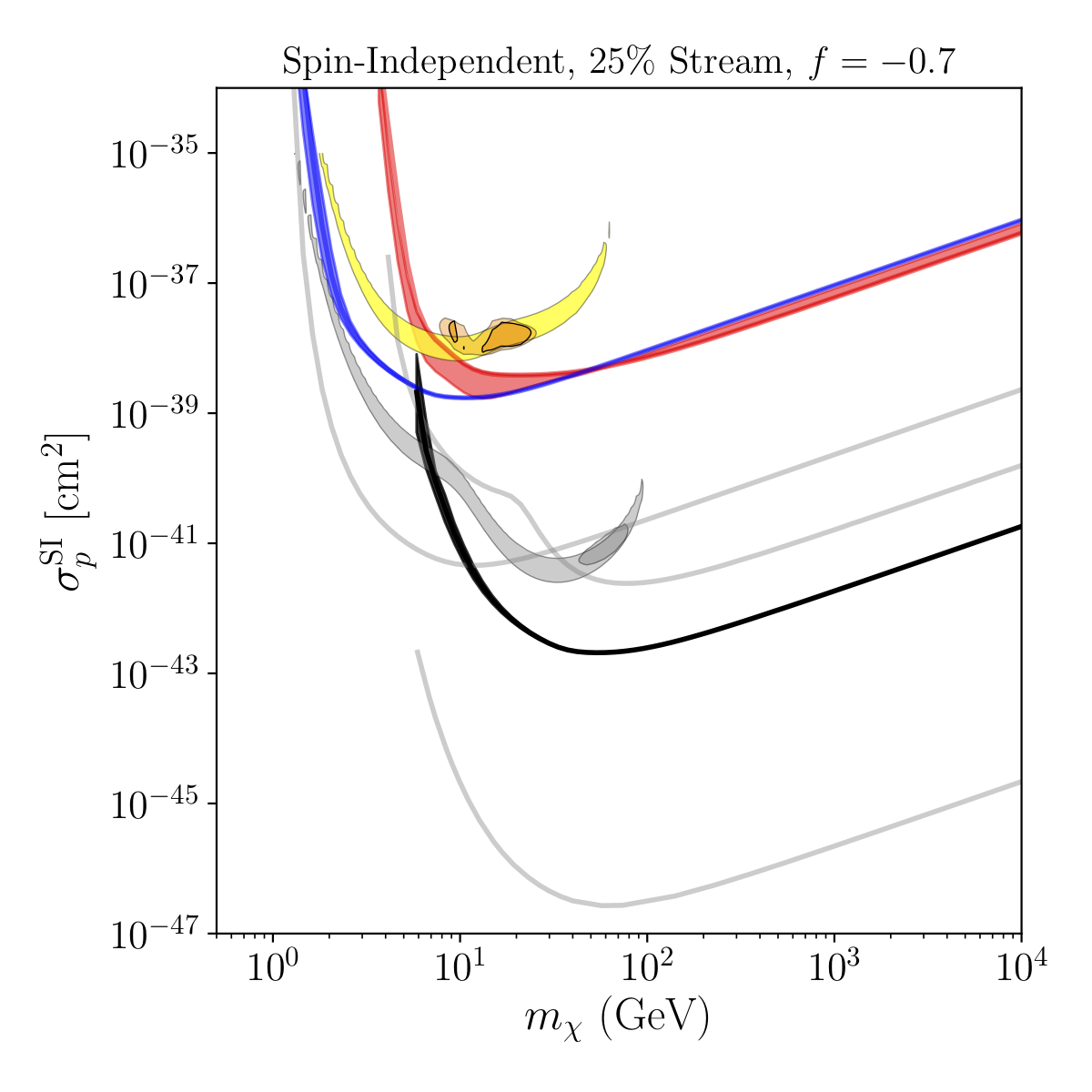}
\caption{90\%CL upper limits from \xenon~(black), \cdms~(blue) and \cosine~(red) on the spin-independent dark matter-proton scattering cross section $\sigma_p^{\rm SI}$ assuming the S1 stream is 100\% of the local dark matter density and an isospin parameter of $f = +1$ (left), 25\% of the local density and $f=+1$ (center), and $25\%$ of the density and $f=-0.7$ (right). The shaded areas in the respective limits indicate the $1\sigma$ uncertainties in the S1 stream velocity distribution variation.
The $2\sigma$ regions around the best-fit to \dama~data are shown in orange for fits to the reported modulation amplitudes (with the lighter shaded region indicating the $1\sigma$ variation of the S1 stream velocity distribution), and in yellow for fits to the reported annual modulation rates. 
Corresponding regions assuming the SHM with $f = +1$ are shown in grey shading.  
\label{fig:S1_SI} }
\end{figure}

We next consider the effect of the S1 stream, treating its contribution to the local dark matter density as a free parameter. To demonstrate the effect of the new velocity profile on the direct detection experiments, on the right panel of Figure~\ref{fig:DAMA_bestfit_mod} we plot the annual modulation event rate on top of the \dama~Phase-2 data and show that the S1 stream peaks on nearly the same day as the data for our best fit parameters.
We also show in the left panel of Figure~\ref{fig:S1_SI} the extrapolated limits if the S1 stream was 100\% of the local density, assuming isospin conservation ($f= +1$). For all the limits, we vary stream parameters from the $1\sigma$ limits of Eq.~\eqref{eq:S1parameters}.

As expected from the behavior of the $\eta$ function shown in Figure~\ref{fig:DMdistributions} the exclusion limits for all experiments strengthen somewhat at low dark matter mass and weaken at high mass. Further, the \dama~regions move to lower mass and lower cross section; the best-fit parameter point occurs at either $\sim 26$~GeV or $\sim 7$~GeV, depending on the stream velocity parameters. However, the $\chi^2/{\rm d.o.f.}$ for these parameter points is between $8$ and $15$ when fit to the binned recoil spectrum (counting the stream parameters as additional degrees of freedom). This indicates that the stream, by itself, is not a particularly good fit to the measured \dama~recoil spectrum (though it is still statistically preferred over the no-signal hypothesis). An example of one of these best-fit parameters for 100\% stream density and $f=+1$ is shown in the right panel of Figure~\ref{fig:DAMA_bestfit}, while the modulation fit is shown in Figure~\ref{fig:DAMA_bestfit_mod}. Critically, in the latter fit, it is apparent that the phase shift of the stream is largely consistent with the observations, with the peak modulation occurring at t = 151 days. 

Of course, it is not reasonable to expect that the S1 stream will constitute 100\% of the local dark matter density. To display limits in the multi-parameter space of dark matter mass $m_\chi$, cross section $\sigma_p^{\rm SI}$, isospin-violating parameter $f$, and stream fraction, we take two approaches. First, as an example, we show in the center and right panels of Figure~\ref{fig:S1_SI} the extrapolated limits and best-fit \dama~regions assuming the stream is 25\% of the local density -- this assumption is perhaps on the high end of the ${\cal O}(10\%)$ estimate, but not implausible. 
Again, we show limits for two choices of the isospin parameter: $f=+1$ and $f=-0.7$. The former has a minimum $\chi^2/{\rm d.o.f.}= 3.9$, while the latter has a minimum goodness-of-fit of 6.7.
Again, we see that the stream can shift the \dama~best-fit to lower masses and therefore bring it closer to the edge of the various exclusion limits, but in the representative parameter choices, the \dama~region continues to be ruled out.

\begin{figure}[t]
\includegraphics[width=0.4\columnwidth]{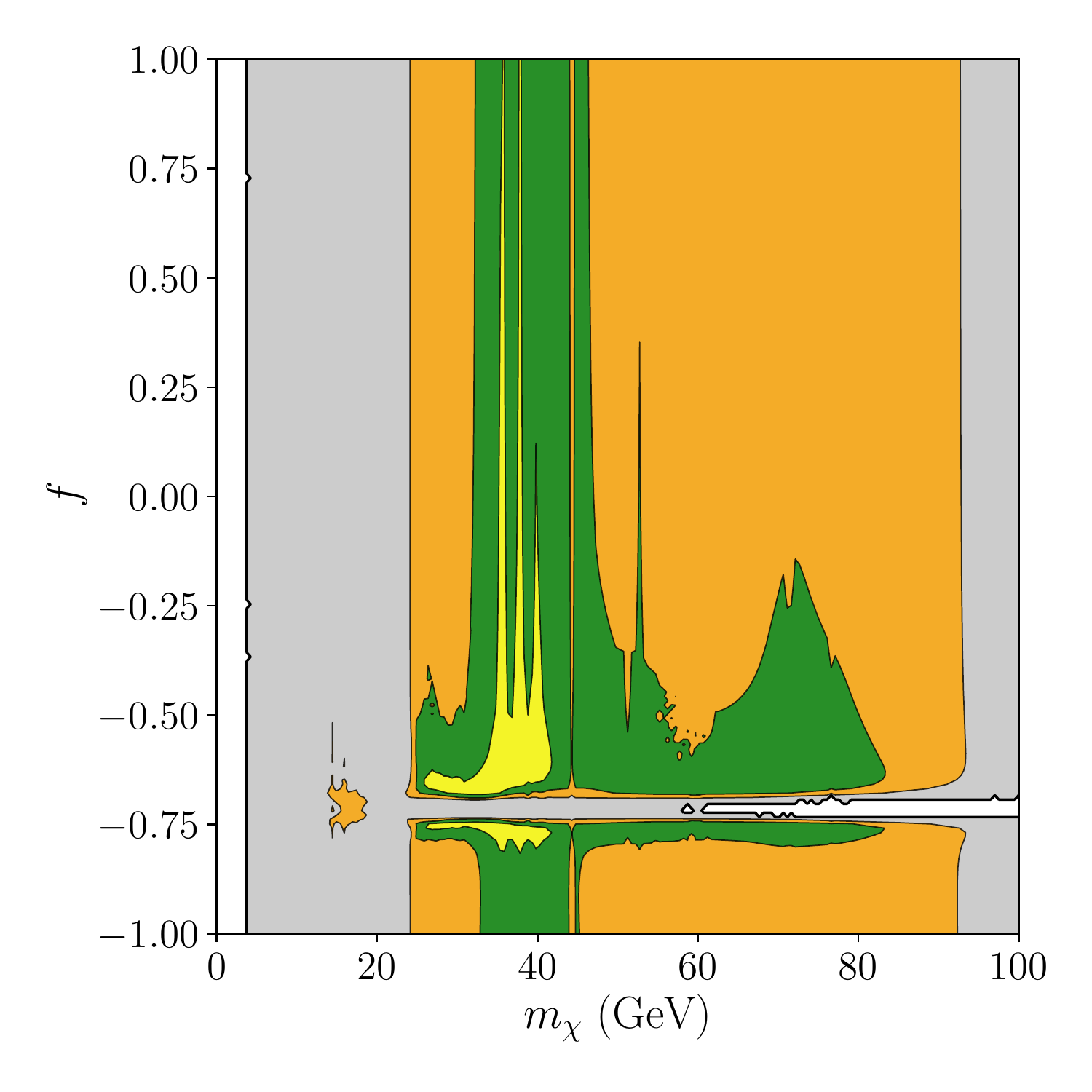}
\includegraphics[width=0.4\columnwidth]{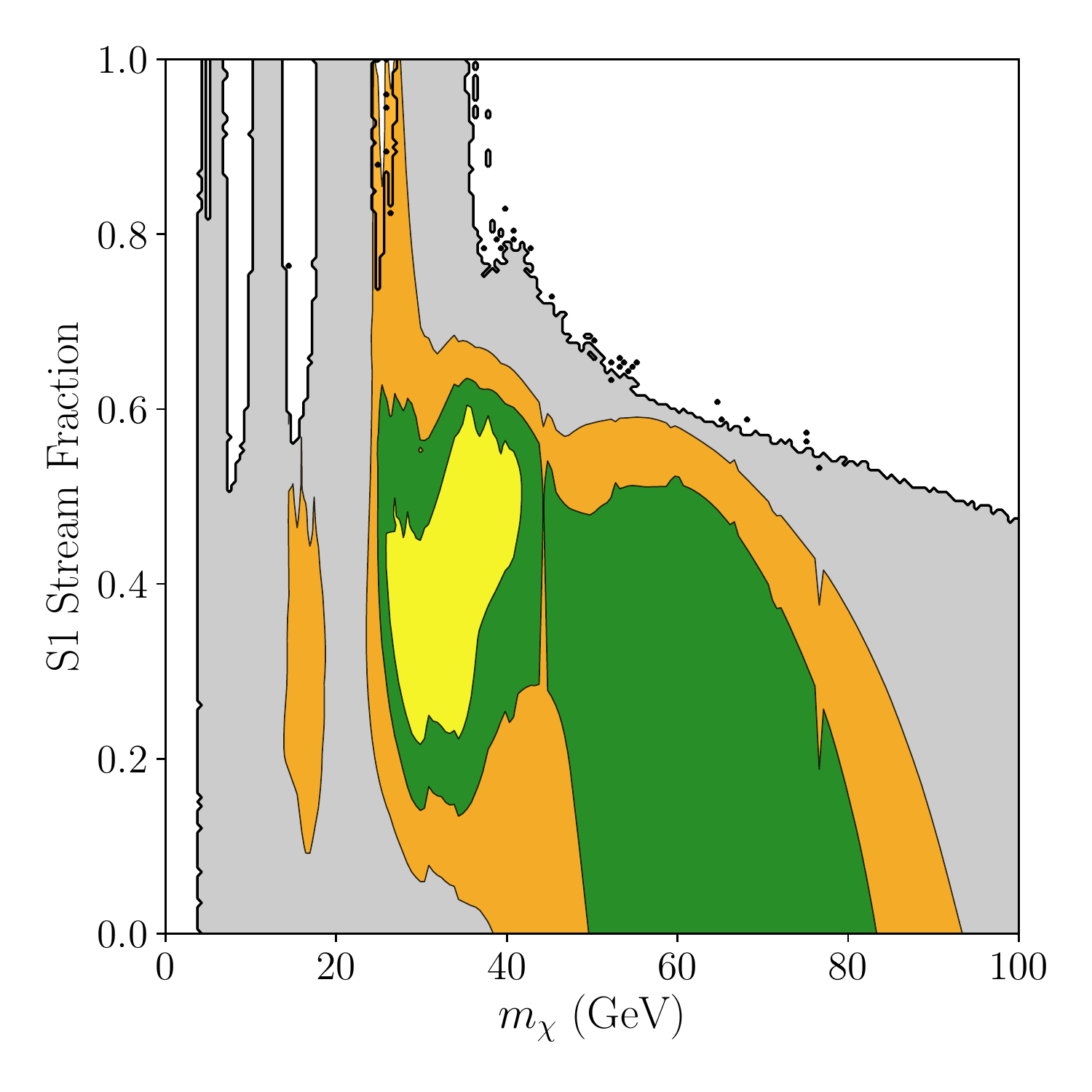}
\caption{Plots showing the $1$, $2$, and $3\sigma$ regions (yellow, green, orange) around the best fit to the \dama~data, profiling over the mass, the proton scattering cross section $\sigma_{p}^{\rm SI}$, stream density fraction, and isospin violating parameter $f$, displayed as a function of mass $m_\chi$ and $f$ (left panel) and $m_\chi$ and stream fraction (right panel). Regions in grey correspond to a parameter point where the cross section which is $3\sigma$ away from the best fit point is excluded by \xenon~bounds (assuming a given mass and $f$/stream fraction).
\label{fig:DAMA_best} }
\end{figure}

However, this set of plots does not completely guarantee that there is not some combination of dark matter mass, cross section, isospin-violation, and stream fraction that would not allow the \dama~region to evade the existing limits. We therefore show in Figure~\ref{fig:DAMA_best} two scans: the first over dark matter mass $m_\chi$ and $f$ (left panel) and the second over $m_\chi$ and the fraction of the local density in the S1 stream (right panel). 
As before, we select stream parameterizations within the $1\sigma$ errors of Eq.~\eqref{eq:S1parameters}. In these plots we display the regions of parameter space that are within $1\sigma$, $2\sigma$, and $3\sigma$ of the best-fit. The best fit point corresponds to $m_\chi = 38$~GeV, $f = -0.60$, $\sigma_p^{\rm SI} = 2.5\times 10^{-39}$~cm$^2$, and a stream fraction of $52\%$, with a $\chi^2/{\rm d.o.f.} = 0.79$. A slightly worse fit ($\chi^2/{\rm d.o.f} = 1.2$) exists at $m_\chi = 27$~GeV, $f = -0.67$, $\sigma_p^{\rm SI} = 8.4\times 10^{-39}$~cm$^2$, and a stream fraction of $34\%$. 
We show the recoil spectrum for this latter point in Figure~\ref{fig:DAMA_bestfit}.

From this analysis, we can see that the addition of the stream allows for a markedly improved fit to the observed spectrum, for reasonable values of the stream density. We can therefore conclude that the addition of streams can noticeably alter the expected spectrum of dark matter recoils in direct detection, when compared to the SHM or smooth halo models predicted from simulation. This is perhaps interesting in light of the \dama~Phase-2 results, which have been noted \cite{Kelso:2013gda,Baum:2018ekm} to be poor fits to low-mass dark matter scattering, due to the observed recoil spectrum at low energies. As seen, even relatively small admixtures of dark matter streams such as S1 can improve the quality of fit.

However, in Figure~\ref{fig:DAMA_best}, along with the fits to the \dama~results, we also show shaded regions which indicate those points where the scattering cross section that corresponds to the minimum $\chi^2$ fit to the \dama~data is itself ruled out by one of the other direct detection experiments, typically \xenon, hence for brevity and simplification of our results we only show the exclusion by \xenon. As can also be seen, for all choices of spin-independent elastic scattering, the \dama~modulation is ruled out for all possible S1 stream parameters (for cross sections up to $3\sigma$ from the best-fit point), with the exception of a very small region around $m_\chi \sim 30$~GeV and stream fractions of more than 80\%, which is marginally within the $3\sigma$ contour around the global best-fit point. Note that no equivalent $3\sigma$ region appears in the scan over mass and $f$. This is due to the fact that for fixed $m_\chi = 30$ GeV and fixed isospin parameter $f$, there is a stream fraction that gives the best fit. We find that this best-fitting stream fraction is never $>0.8$, thus every point with $m_\chi =30$ GeV on the left-handed plot is ruled out by Xenon. On the other hand, forcing the stream fraction to be 90\%, gives a best value for $f$ that fits the mass/stream fraction requirement. This combination of mass/fraction/$f$ is not ruled out, which is illustrated on the right panel. However, since the left panel is profiling over the stream fraction, the allowed point is missed. This indicates that a non-excluded region requires some of the model parameters to be perturbed from values that would minimize the $\chi^2$ fit at fixed $m_\chi$ and $f$.

\subsection{Spin-Dependent Elastic Scattering}

\begin{figure}[t]
\includegraphics[width=0.325\columnwidth]{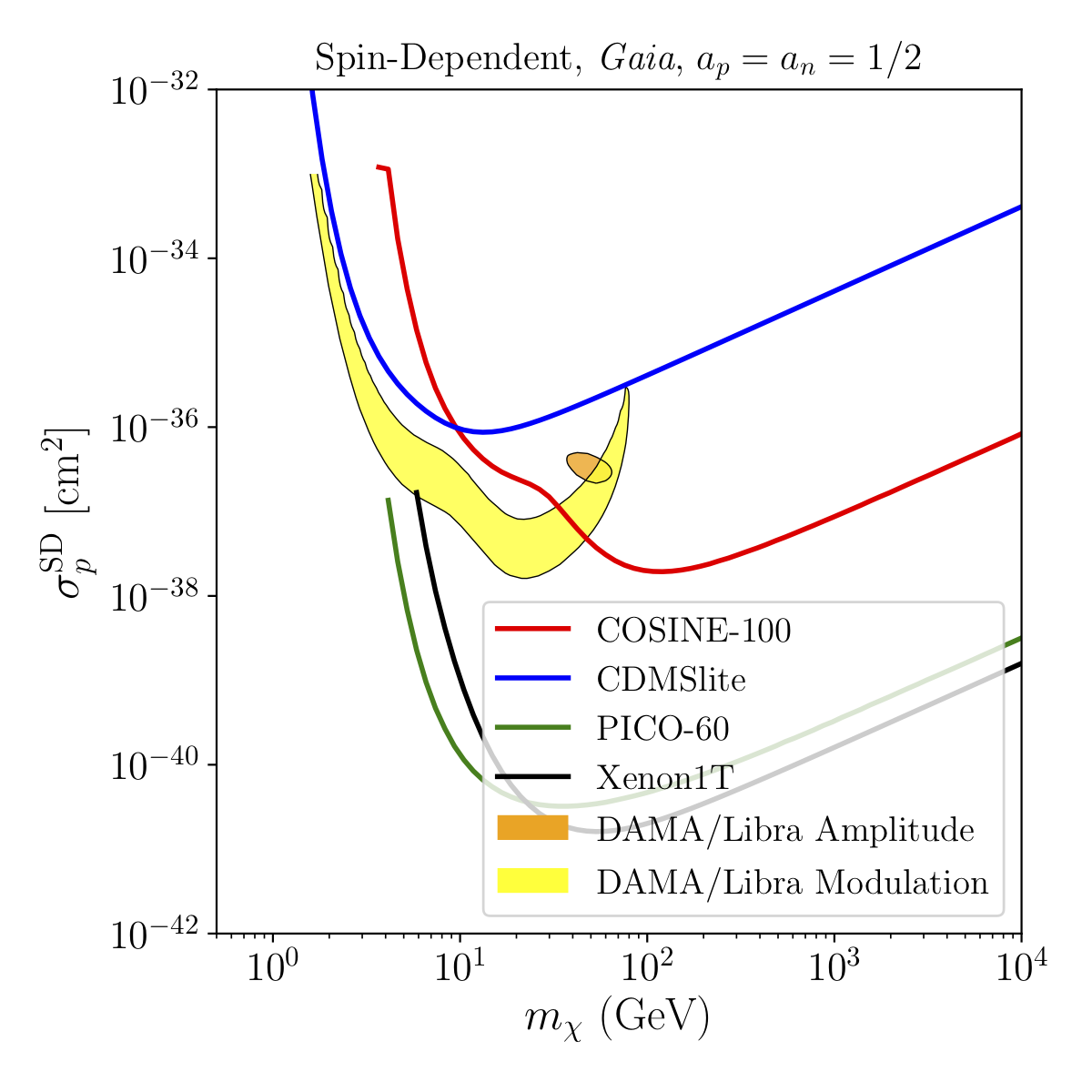}
\includegraphics[width=0.325\columnwidth]{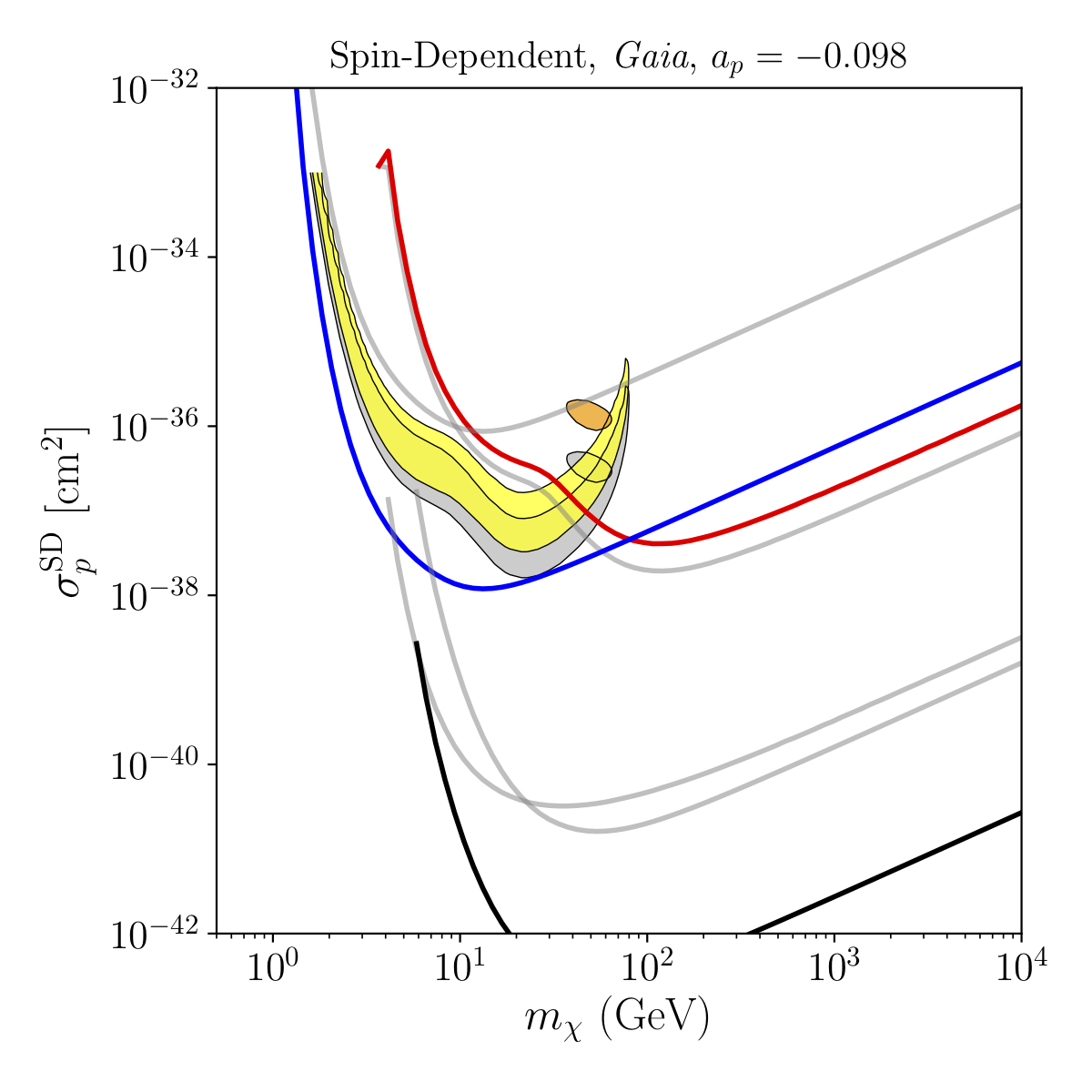}
\includegraphics[width=0.325\columnwidth]{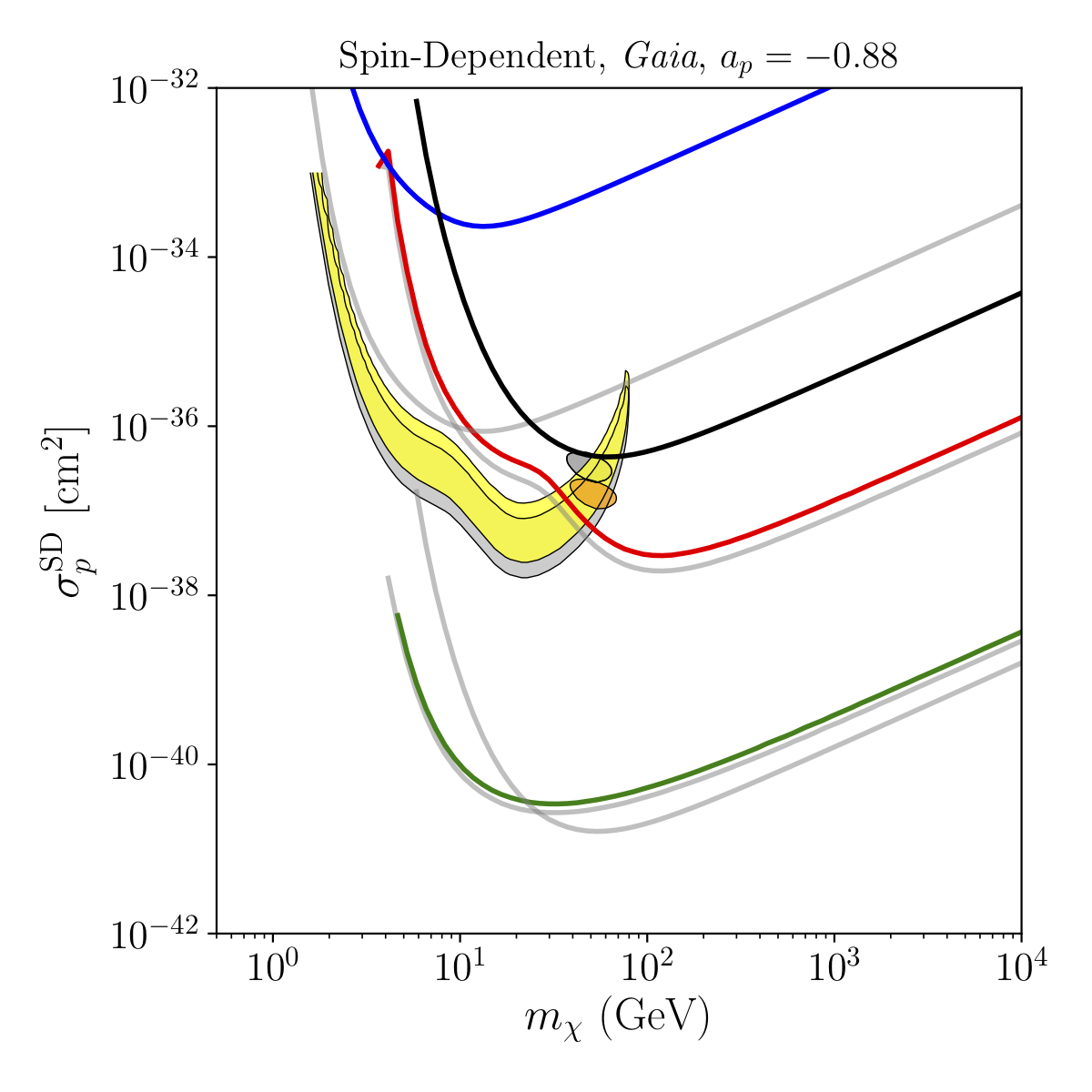}
\caption{90\%CL upper limits from \xenon~(black), \cdms~(blue) and \cosine~(red) on the spin-dependent dark matter-proton scattering cross section $\sigma_p^{\rm SD}$ assuming the \gaia~halo model for isospin-conserving interactions $a_p = a_n = 1/2$ (left), $a_p = -0.098,~a_n = +0.902$ (minimum interaction with fluorine, center) and $a_p = -0.88,~a_n = +0.12$ (minimum interaction with xenon, right). Best-fit $2\sigma$ regions to the \dama~data are shown in orange for fits to the reported modulation amplitudes, and in yellow for fits to the reported annual modulation rates. Equivalent isospin-conserving limits assuming the \gaia~halo model are shown in shaded grey in the right two plots for comparison purposes. \label{fig:SDnostream_isospinconserving} }
\end{figure}

We now turn to spin-dependent elastic scattering. In Figure~\ref{fig:SDnostream_isospinconserving} we show the extrapolated limits on the proton scattering cross section $\sigma_p^{\rm SD}$, assuming the isospin-conserving couplings $a_p = a_n = 1/2$ and the \gaia~halo model without the S1 stream. We also display two isospin-violating scenarios, one where $a_p = -0.098$ and $a_n = +0.902$ (chosen to minimize the scattering off fluorine and thus relax the \pico~bound,) and the second assuming $a_p = -0.88$ and $a_n = +0.12$ (chosen to minimize the \xenon~limit).\footnote{Here we assume the loop-induced mixing of the somewhat-misnamed ``neutron''  and ``proton'' couplings is $\delta a = -0.2$, per the discussion around Eq.~\eqref{eq:SDmixing}. Varying the value of $\delta a$ will change the values of $a_p$ and $a_n$ corresponding to exact cancellation, but will not qualitatively change the results.} As with the spin-independent scattering, the \dama~regions continue to be excluded assuming a smooth halo model as extracted from the \gaia~data. Assuming $a_p = a_n = 1/2$, we find a best fit point of $m_\chi = 48$~GeV, $\sigma_p^{\rm SD} = 3.6 \times 10^{-37}$~cm$^2$ and $\chi^2/{\rm do.f.} = 1.9$. Profiling over $a_p$, we find little improvement in the halo-only model, with the best fit occurring for $a_p = -0.67$, but at essentially the same mass, cross-section, and $\chi^2$ value.\footnote{The best-fit mass for spin-dependent scattering does not shift significantly as $a_p$ is varied because the ratio of $\langle S_n\rangle/\langle S_p \rangle$ for sodium is very similar to that of iodine. Thus, one can not ``turn off'' scattering against the heavier iodine in favor of sodium through a choice of $a_p$, as can be done in spin-independent scattering through a choice of $f$.}

\begin{figure}[t]
\includegraphics[width=0.325\columnwidth]{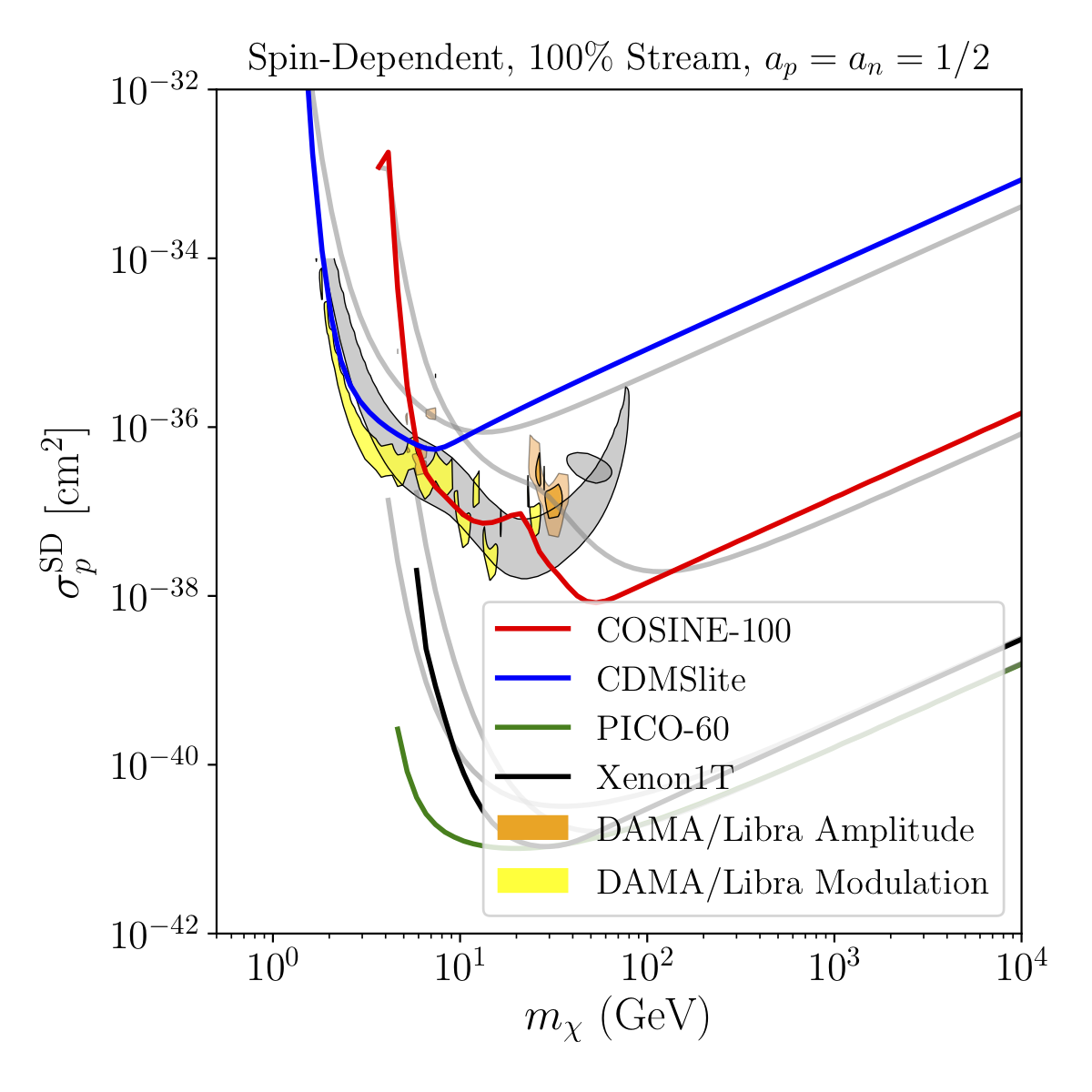}
\includegraphics[width=0.325\columnwidth]{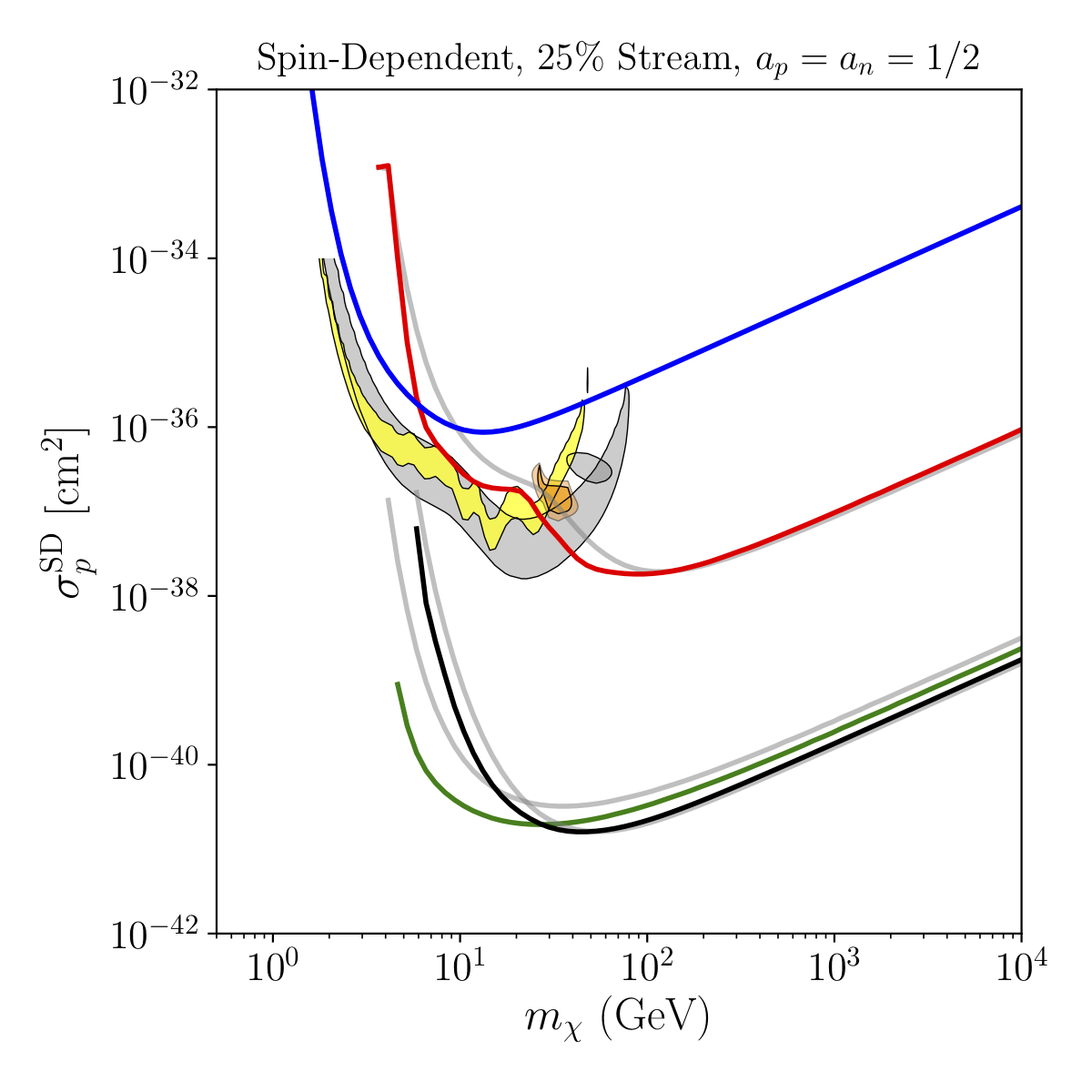}
\includegraphics[width=0.325\columnwidth]{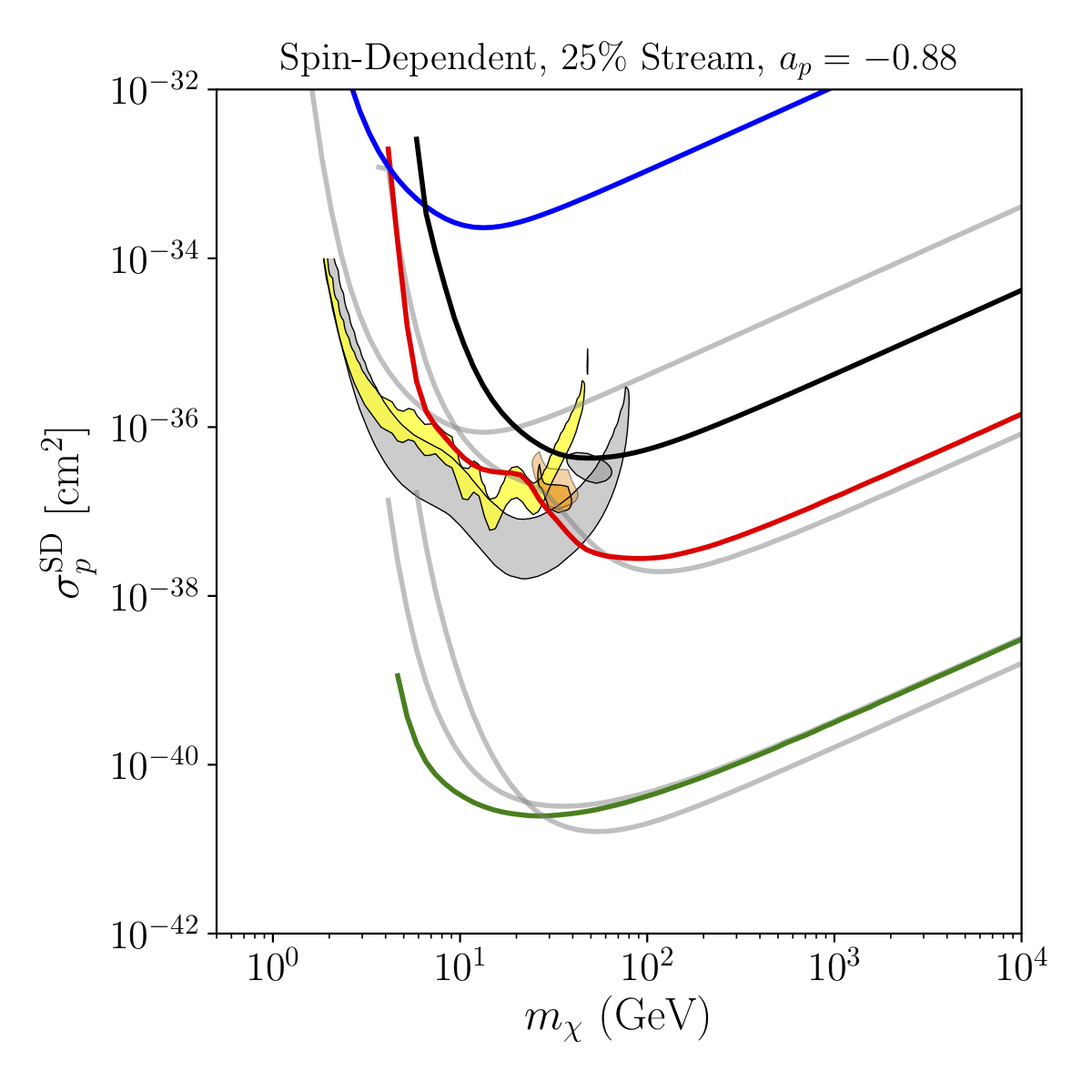}
\caption{90\%CL upper limits from \xenon~(black), \cdms~(blue) and \cosine~(red) on the spin-dependent dark matter-proton scattering cross section $\sigma_p^{\rm SI}$ assuming the S1 stream is 100\% of the local dark matter density and an isospin parameter of $a_p =a_n = +1/2$ (left), 25\% of the local density and $a_p=a_n=+1/2$ (center), and $25\%$ of the density and $a_p=-0.88$ (right).
The $2\sigma$ regions around the best-fit to \dama~data are shown in orange for fits to the reported modulation amplitudes (with the lighter shaded region indicating the $1\sigma$ variation of the S1 stream velocity distribution), and in yellow for fits to the reported annual modulation rates. 
Corresponding regions assuming the \gaia~halo model with $a_p =a_n= +1/2$ are shown in grey shading.
\label{fig:S1_SD} }
\end{figure}

We can now introduce the S1 stream. As in the spin-independent case, we demonstrate the effect of the S1 stream in Figure~\ref{fig:S1_SD}, showing the extrapolated results assuming 100\% of the local density comes from S1, and isospin-conserving couplings. We also show the more reasonable 25\% stream density, with $a_p = a_n =1/2$ and $a_p = -0.88$ (chosen to minimize the xenon couplings). The minimum $\chi^2/{\rm d.o.f.}$ under these assumptions is 6.9 for the 100\% stream density, and 1.4 (1.45) for the 25\% density with $a_p = 1/2$ ($a_p = -0.88$). Thus, as with elastic spin-independent scattering, the stream by itself does not improve the fit to the observed recoil spectrum, but a reasonable admixture of the smooth halo and the stream results in a much improved goodness-of-fit parameter. As the resulting recoil spectrum and annual modulation strongly resembles that of the spin-independent scattering, we omit the spin-dependent equivalents of Figures~\ref{fig:DAMA_bestfit} and \ref{fig:DAMA_bestfit_mod}.

\begin{figure}[t]
\includegraphics[width=0.45\columnwidth]{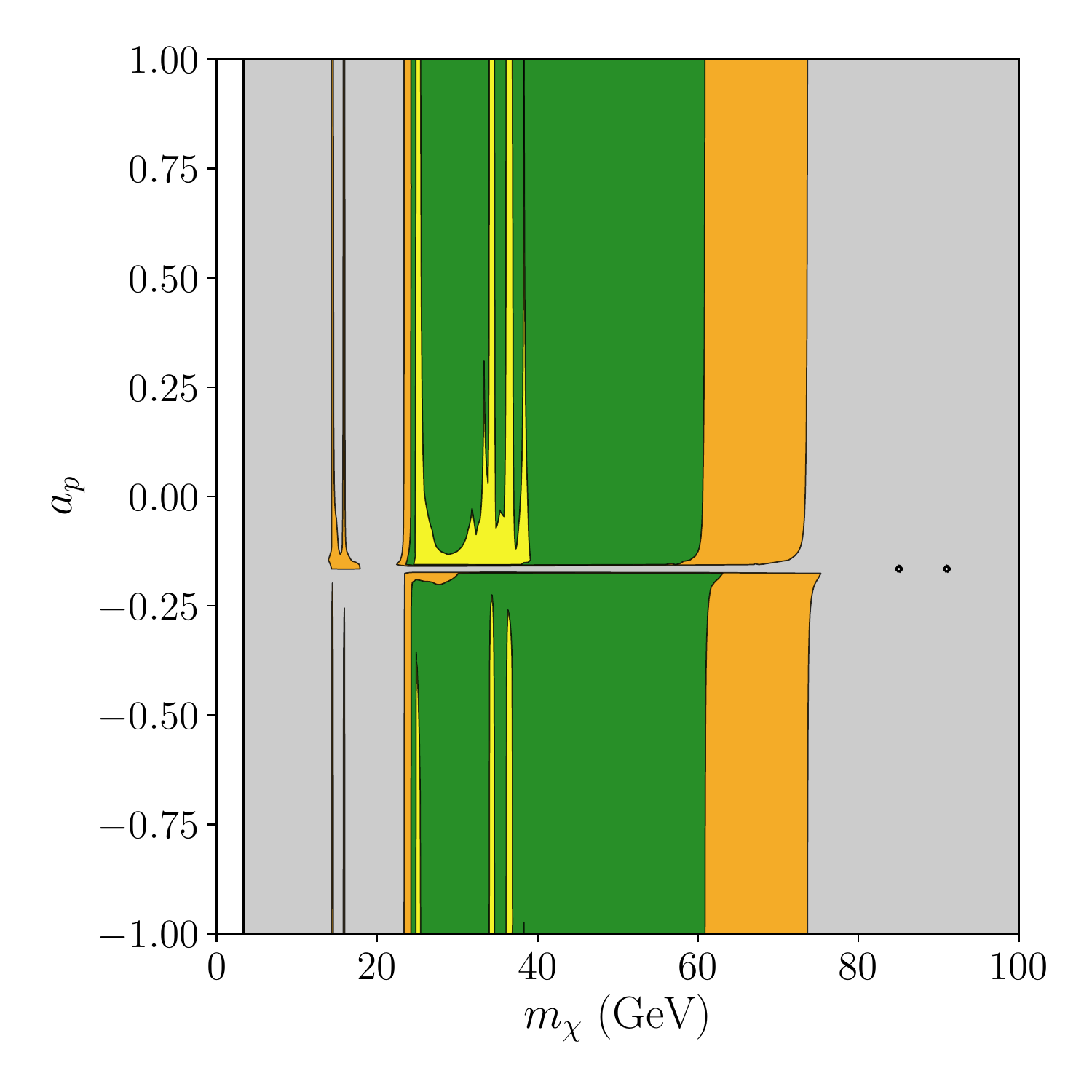}
\includegraphics[width=0.45\columnwidth]{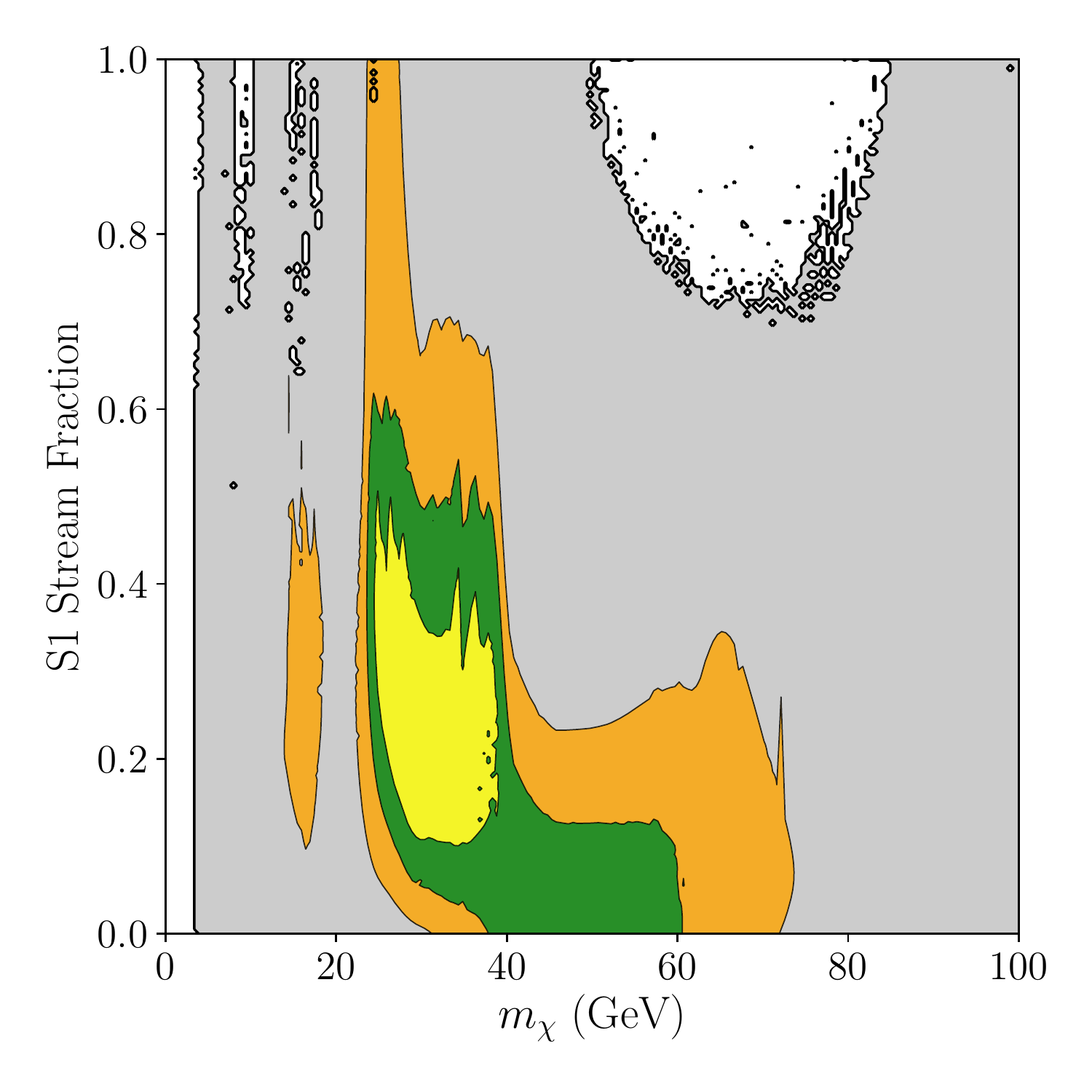}
\caption{Plots showing the $1$, $2$, and $3\sigma$ regions (yellow, green, orange) around the best fit to the \dama~data, varying the mass, the spin-dependent proton scattering cross section $\sigma_{p}^{\rm SD}$, stream density fraction, and isospin violating parameter $a_p$, displayed as a function of mass $m_\chi$ and $a_p$ (left panel) and $m_\chi$ and stream fraction (right panel). Grayed-out regions correspond to a parameter point where the cross section which is $3\sigma$ away from the best fit point which is excluded by \xenon~bounds (assuming a given mass and $a_p$/stream fraction).
\label{fig:DAMA_best_SD} }
\end{figure}

 As with spin-independent scattering, given the breadth of possible parameter space, we cannot ensure that all regions of \dama~parameter space are excluded by other null results by taking slices through $a_p$ space. We again perform two scans over mass and $a_p$ space, fitting to the \dama~amplitude data. In the first scan, we assume the best fit stream fraction and $\sigma_p^{\rm SD}$. For the second, we scan over mass and stream fraction, for the best fit $a_p$ and cross section. For each parameter point, we check to see if the required cross section for a $3\sigma$ fit to the \dama~data is ruled out by any other experiments at 90\%CL. The results are shown in Figure~\ref{fig:DAMA_best_SD}. As can be seen, no region of parameter space survives this test.

It should also be noted that the modulation data alone can have good fits in regions of $m_\chi-\sigma_p^{\rm SD}$ parameter space which are {\it not} excluded by any null result. This occurs without the S1 stream for a small region of low-mass parameter space with $a_p = -0.88$ (right panel, Figure~\ref{fig:SDnostream_isospinconserving}), and becomes more pronounced as the stream is added (Figure~\ref{fig:S1_SD}). These regions are not within in the $2\sigma$ good-fit regions to the amplitude data. So, {\it if} the \dama~signal from the modulation data alone were to be interpreted as a signal of dark matter scattering through an elastic spin-dependent interaction, then the observed energy recoil spectrum as measured by \dama~must be in significant tension with the actual recoil spectrum. 

For example, consider a parameter point which is a good fit to the modulation data and is not ruled out by other experiments: $a_p = 1/2$, a stream fraction of 25\%, $m_\chi = 4$~GeV and $\sigma_p^{\rm SD} = 4\times 10^{-36}$~cm$^2$. This parameter point results in an improvement of $\Delta \chi^2 \sim 24$ over the null hypothesis for the modulation data. However, this point is only a $\chi^2/{\rm d.o.f.} \sim 1.5$ improvement over the null hypothesis in the recoil spectrum data, and would be $\chi^2/{\rm d.o.f.} \sim 30$ away from the best possible fit in the $m_\chi/\sigma_p^{\rm SD}$ plane. Therefore, while including the stream at reasonable density fractions can fit the observed \dama~modulation data while evading all other null constraints, some significant systematic errors would have to be present to reconcile the reported \dama~recoil spectrum with that caused by dark matter.

\subsection{Inelastic Dark Matter}

\begin{figure}[t]
\includegraphics[width=0.45\columnwidth]{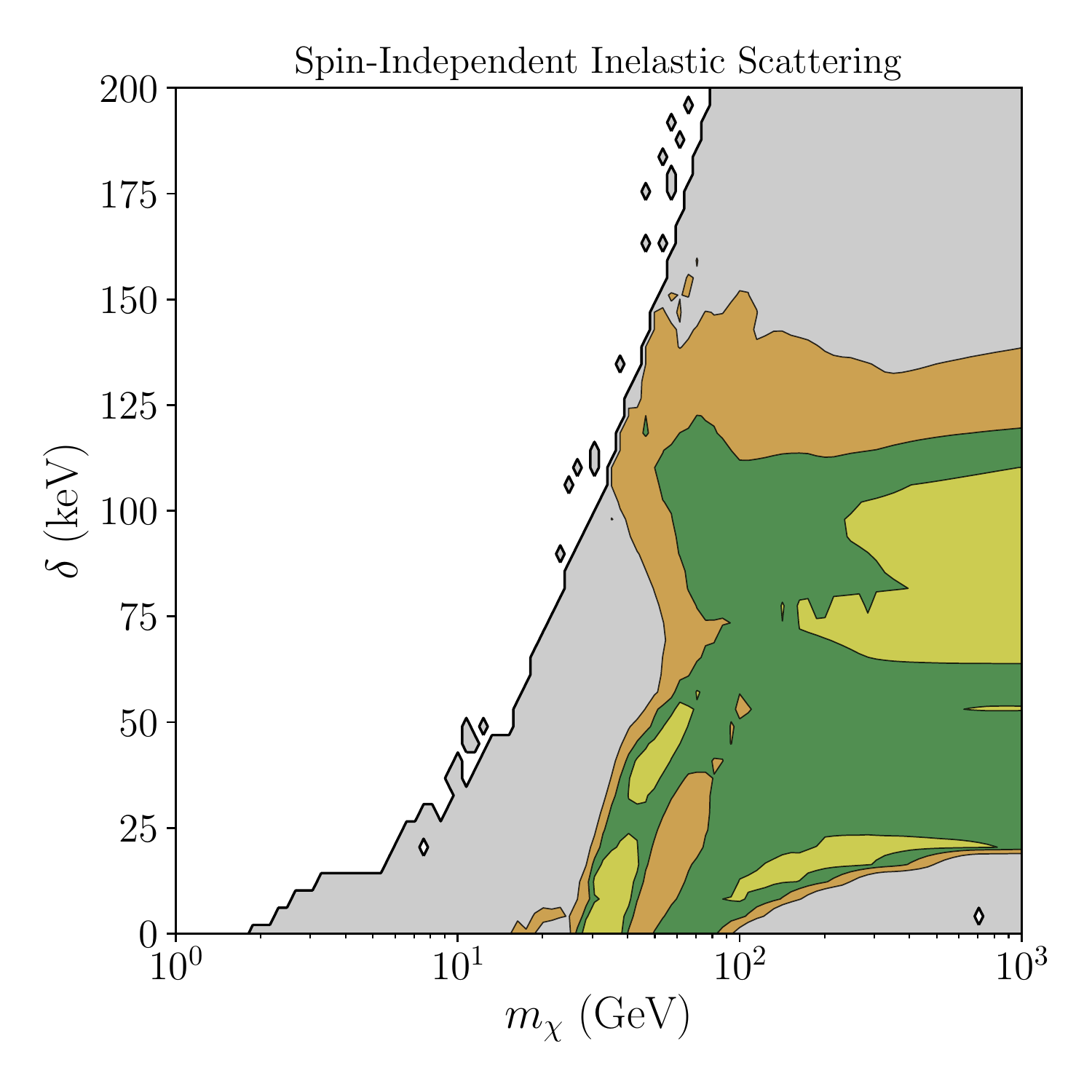}
\includegraphics[width=0.45\columnwidth]{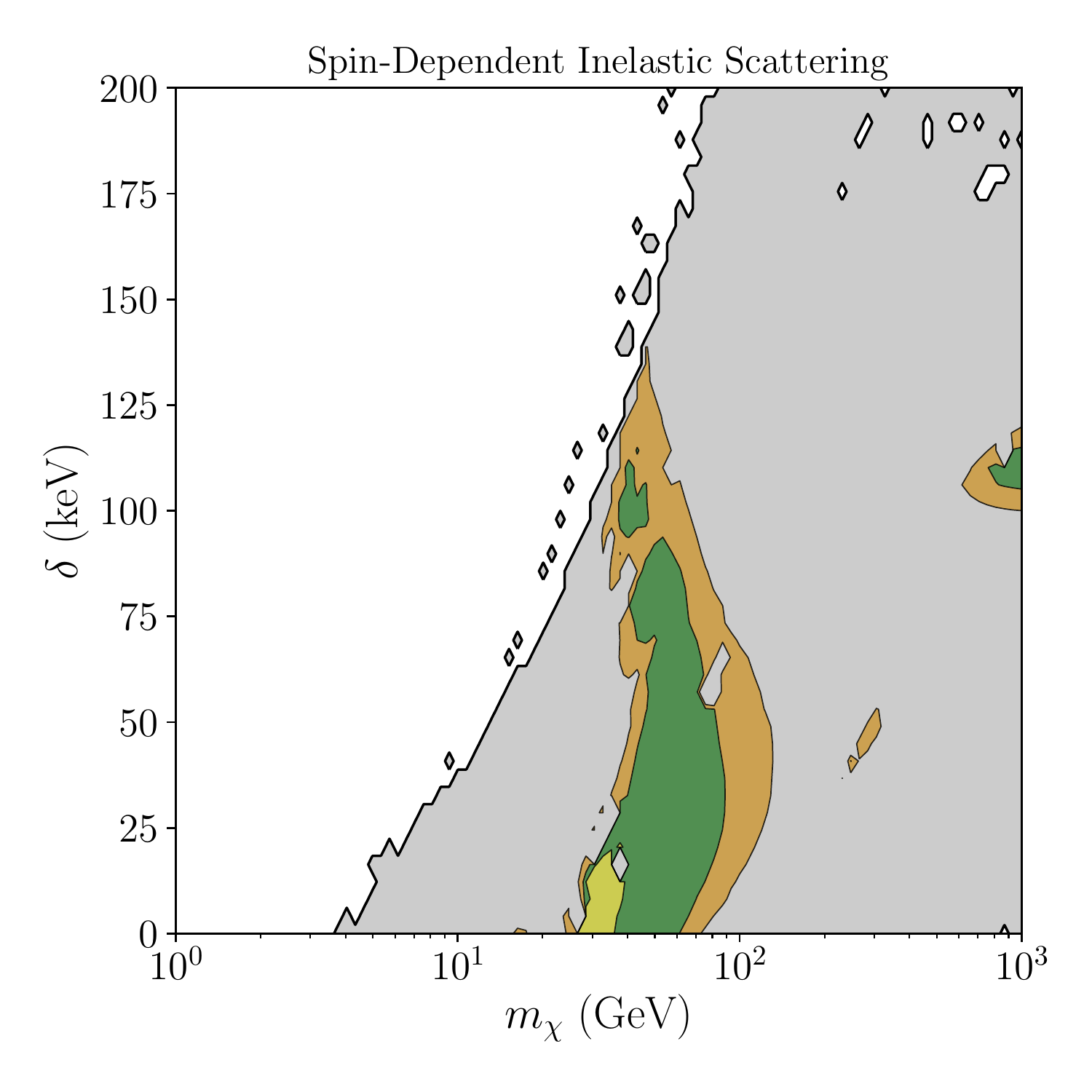}
\caption{Plots showing the $1$, $2$, and $3\sigma$ regions (yellow, green, orange) around the best fit to the \dama~data, varying the mass and mass splitting $\delta$ for spin-independent scattering (left panel) and spin-dependent scattering (right panel). The grayed-out regions correspond to a parameter point where the cross section which is $3\sigma$ away from the best fit point is excluded by \xenon~bounds. 
\label{fig:DAMA_best_iDM} }
\end{figure}

Having reanalyzed the elastic scattering constraints in the presence of the S1 stream, we now turn to the constraints assuming the inelastic dark matter scenario. Here, scattering proceeds through a ground dark matter state which up-scatters into an excited state, with mass difference $\delta$. In elastic scattering, the dark matter mass sets the recoil spectrum through both the implicit dependence of $v_{\rm min}$ on $m_\chi$ and the explicit mass dependence in Eq.~\eqref{eq:dRdER}. Inelastic scattering on the other hand allows an additional handle by modifying the minimum dark matter velocity required for a scattering of recoil energy $E_R$, as described in Eq.~\eqref{eq:idm_vmin}. 
However one new degree of freedom: the mass splitting $\delta$ is added to the model parameters (cross-section, mass, isospin-violating parameter $f$ and stream density fraction) and to study the viable parameter space we must scan over these.
Since we are dealing with a multi-dimensional parameter space, similar to the elastic scattering case above, we test whether the \dama~best fit cross-sections are excluded by the other experiments, when varying the extra inelastic degree of freedom.
Hence in Figure~\ref{fig:DAMA_best_iDM} we fit the \dama~amplitude data, scanning over dark matter mass $m_{\chi}$ and mass spitting $\delta$ (for both spin-independent and spin-dependent scattering). We show the $1$, $2$ and $3\sigma$ regions around the best fit point, portrayed in yellow, green and orange respectively. 
For spin-independent scattering (left panel), the best-fit point for spin-independent scattering is $m_\chi = 230$~GeV, $\delta = 20$~keV,  $f=-0.79$, $\sigma_{p}^{SI} = 7.7\times 10^{-37}$~cm$^2$ with a stream density fraction of 47\% and $\chi^2/{\rm d.o.f.} = 0.9$. There are nearly as good fits ($\chi^2/{\rm d.o.f.} \sim 1$) at lower mass points as well. However, for the entire range of $m_\chi$ and $\delta$, the best fit parameter points are ruled out by other experimental results, in particular \xenon~(as shown by the grey shaded region).

For spin-dependent scattering, the best-fit point occurs when the inelastic scattering is turned off ($\delta = 0$), and $m_\chi = 28.5$~GeV, $a_p = -0.15$, $\sigma_{p}^{SD} = 9\times 10^{-34}$~cm$^2$, and benchmark stream density of 24\%. An island of similarly good fits occurs around these parameters for $\delta \lesssim 20$~keV. 
As seen in Figure~\ref{fig:DAMA_best_iDM}, the \dama~$3\sigma$ and \xenon~exclusion regions nearly coincide for $m_\chi \sim 35$~GeV. This is due to the fact that, as we move away from the $3\sigma$ region of parameter space, the worsening fit to the \dama~data drives the cross section of the best possible fit lower and hence closer to the \xenon~limit.

\section{Discussion and Conclusions}
The dark matter velocity distribution in the local Galactic neighborhood has long been postulated to follow a Gaussian Standard Halo Model. Astrophysical data however, points to the presence of new dark matter substructure that contributes significantly to the velocity distribution on top of the smooth background halo model. Recently, the \gaia~space telescope observed a number of dark matter streams and clumps moving at high velocities near the Solar System. 
One of these streams, S1, is moving on a low inclination, rapidly counter-rotating orbit through the local Solar neighborhood. Due to its trajectory, the relative velocity between dark matter within S1 and the Earth is increased, enhancing the modulation rate of scattering events in experiments on the Earth.

In this work we reanalyzed the limits from current direct detection experiments, including not only the S1 stream, but also data driven local velocity distributions derived from the \gaia~low and intermediate metallicity stellar population.
We considered various scenarios for dark matter scattering, including elastic and inelastic scattering together with spin-independent and spin-dependent dark matter-nucleus interactions in both scattering circumstances. In each case, we also considered various isospin violating couplings, which could result in the relative suppression or enhancement of scattering rates between different direct detection experiments.

We directly compared the SHM, the \gaia~derived velocity distributions, and the S1 stream and illustrated the effect each of these distributions has on the number of dark matter particles with sufficient kinetic energy to initiate a scattering event above threshold for a direct detection experiment.
We showed that the S1 stream has a distinctive recoil spectrum which could be observed in direct detection experiments as compared to the SHM and the \gaia~distribution, while maintaining a peak modulation date close to the SHM expectation.

We then studied how the existing experimental limits placed on particle physics parameters ($m_{\chi}$ and $\sigma$) change in the presence of S1 and the \gaia~distribution. To do this we first focused on the \dama~experiment which claimed to have seen a modulation signal of dark matter recoils at 13$\sigma$ CL. We fit the \dama~Phase-2 amplitude and annual modulation data for different parameters in the dark matter theory and astrophysics space (e.g. interaction type, isospin ratio, stream fraction, etc). 
In addition to \dama~we considered those experiments which report a null signal for dark matter and place the strongest current limits on spin-dependent or spin-independent scattering (\cdms, \xenon~and \pico). We also took into account the recent limits from the \cosine~experiment which use the same target material as \dama.

As S1 increases the modulation rate and introduces new structures in the recoil spectrum, we find that even for relatively modest contributions of the stream to the local density ($\sim 20-30\%$), the statistical fit to the \dama~data is dramatically improved for both SI and SD interactions with different isospin ratios. 
We find that fits including S1 prefer generally lower dark matter masses compared to the SHM or the \gaia~model for the smooth halo. However, despite these changes in the best fit regions, we find that the null experiments continue to exclude the \dama~allowed region even in the presence of S1 and \gaia~distribution, for our chosen benchmark parameters. 

To ensure we were not missing any crucial model points by simply taking slices in parameter space, we scanned over the stream fraction and isospin factors for the best fitting \dama~dark matter-nucleon cross-section and compared with that for \xenon. We found that for either elastic or inelastic scattering (both SI and SD), the \dama~(modulation amplitude) best fit point continues to be excluded by \xenon, with the exception of spin-independent elastic isospin-violating scattering with large contributions ($>80\%$) to the local density from the stream. We can therefore conclude that the anomalous \dama~results continue to be excluded by the null results from other experiments unless one (or both) of the following conditions are met:
\begin{enumerate}
\item The S1 stream (or a collection of other streams with similar kinematic parameters) composes the vast majority of the local density of dark matter (greater than $\sim 80\%$). This would not be the standard expectation, but would allow for a recoil spectrum matching that of \dama~while not being excluded at the 90\%CL by any other experiment.

\item The dark matter signal is present, but the dark matter recoil spectrum from the modulating signal is in significant ($\gtrsim 3\sigma$) tension with the \dama~results. As we have shown, the best-fits from the yearly modulation data alone allows for low-mass fits when including the S1 stream at reasonable density that are not excluded by any other experiment. However, while the overall modulation rate matches the data, the resulting recoil spectrum would be in significant tension with \dama. 
\end{enumerate}
Measuring the contribution of the S1 stream to the local density would go far in testing the first option, in addition to being extremely interesting in its own right as a probe of dark matter substructure and galaxy evolution, as well as the impact on other direct detection experiments of a non-SHM recoil spectrum. The second possibility will be tested by increased data from NaI experiments such as \cosine~or \textsc{SABRE} \cite{Froborg:2016ova}.
We note also that it is highly likely that the effects of S1 or similar streams would also affect lower mass dark matter scattering off electrons, we leave this for future considerations.

\bigskip
\bigskip

\textbf{Acknowledgements.} We are grateful to Vasily Belokurov, Wyn Evans, David Hogg and GyuChul Myeong for valuable discussions and to Lina Necib for helpful comments and providing us with the \gaia~velocity distribution data files. MRB is supported by DOE grant DE-SC0017811. GM and CM are supported by DOE grant DC-SC0012704.  

\appendix
\section{Recasting of Experiment Results \label{app:recast}}

We wish to compare the results of different direct detection experiments while varying the dark matter velocity distribution. This requires recasting the publicly available data to predict the experimental results for distributions besides the Standard Halo Model. In this section, we describe our technique for each experiment, and validate our results for the SHM distribution assuming isospin-conserving spin-independent (SI) scattering when validating \dama~\cite{Bernabei:2018yyw}, \xenon~\cite{Aprile:2018dbl}, \cdms~\cite{Agnese:2017jvy} and \cosine~\cite{Adhikari2018}; and spin-dependent (SD) scattering for \pico~\cite{Amole:2017dex}. 
The nuclear parameters for the isotopes used in the experiments considered this work are listed in Table~\ref{tab:nuc_params}.

\subsection{DAMA/Libra}
\dama~uses a sodium-iodide (NaI) crystal as the detector target. Critically, unlike other direct detection experiments, \dama~does not aim for zero- (or at least very low)-background. Rather, the experiment looks for the annual modulation of scattering events, resulting from the yearly modulation of $f(v)$ as the Earth orbits the Sun. 
For many years now, the \dama~experiment has reported a positive signal, observing a yearly modulation with $12.9\sigma$ significance in the full (Phase-2) data-set \cite{Bernabei:2018yyw}. The phase of the modulation matches the date expected from the Earth's motion through the Milky Way's dark matter halo, peaking at day $\sim150$ after January 1$^{\rm st}$, depending on the range of recoil energies considered.

The \dama~Phase-2 results (corresponding to 1.13 ton$\times$year) \cite{Bernabei:2018yyw} provides the modulation data in two ways. First, they provide the modulation amplitude $S_m$ (fitting to a sinusoidal modulation) as a function of electron recoil energy from 1 to 20~keV. The signal is limited to recoil energies from 1 to 6~keV. 
Secondly, they provide the residual rate of scattering events as a function of time for three recoil energy ranges: 1-3~keV, 2-6~keV, and 1-6~keV. We consider fits to both slicings of the data separately. Importantly, though the modulation of a single dark matter component (i.e., the Galactic halo or the S1 stream) is typically sinusoidal, the sum of two or more components does not need to be.
While the S1 stream has modulation peaks which are nearly in phase with the halo itself (indeed, this coincidence is required for S1 to possibly fit the \dama~signal), the annual modulation data provides important complimentary information which is somewhat obscured by the more finely-binned amplitude results.

To convert between the measured electron recoil energy $E_e$ and the dark matter-induced nuclear recoil energy, we convolute the response function \cite{Baum:2018ekm,Bernabei:2008yh}
\begin{eqnarray}
\phi(E_R,E_e) & = & \frac{1}{\sqrt{2\pi} \sigma(Q E_R)} e^{-(E_e-QE_R)^2/2\sigma(QE_R)^2} \\
\sigma(Q,E_R) & = & \alpha\sqrt{Q E_R} + \beta Q E_R,
\end{eqnarray}
with the differential nuclear recoil distribution $\frac{dR}{dE_{R}}$ and a detector efficiency function $\epsilon$.
Here $\alpha = (0.448 \pm 0.035)~{\rm keV}^{1/2}$ and $\beta = (9.1 \pm 5.1) \times 10^{-3}$. We use quenching factors $Q_{\rm I} = 0.09$ and $Q_{\rm Na} = 0.3$ \cite{Bernabei:1996vj}. 

The electron equivalent recoil distribution is given by 
\begin{equation}
\frac{dR}{dE_{e}} = \int_{0}^{\infty} dE_{R}~ \epsilon~\phi(E_{R}, E_{e}) \frac{dR}{dE_{R}}.
\label{eq:electron_equiv}
\end{equation}
For the purposes of this work we assume $\epsilon = 1$.
The total recoil rate in units of [1/keV/kg/day] is obtained by integrating the electron equivalent recoil distribution per energy interval per time:
\begin{equation}
R_{j} (m_{\chi}, \sigma_{p}) = \frac{1}{\Delta t ~\Delta E_{e}} ~\int_{E_{e}^{\rm min}}^{E_{e}^{\rm max}} \int_{t_{\rm min}}^{t_{\rm max}} ~\frac{dR}{dE_{e}} (m_{\chi}, \sigma_{p})~dE_{e} ~dt,
\end{equation}
where the $E_{e}^{\rm min}$ and $E_{e}^{\rm max}$ are the lower and upper bounds for the energy bin under consideration, while $t_{\rm min}$ and $t_{\rm max}$ are the lower and upper bounds for the time interval, with $\Delta E_{e}$ and $\Delta t$ the energy and time bin widths.
The \dama~ annual modulation data is usually presented in terms of a residual recoil rate defined as $R - \bar{R}$, where $\bar{R}$ is the the mean rate averaged over the total number of time bins in an energy interval.
We perform a $\chi^{2}$ fit to the \dama~Phase-2 residual data using
\begin{equation}
\chi^{2} (m_{\chi}, \sigma_{p}) = \sum_{j} \frac{[ {\cal O}_j - {\cal S}_j(m_{\chi}, \sigma_{p})]^{2}}{\sigma_{j}^{2}},
\label{eq:chi2res}
\end{equation} 
where ${\cal O}_j$ is the observed residual rate in bin $j$, ${\cal S}_j$ is the predicted signal residual rate (dependent on the input dark matter mass and scattering cross-section), and $\sigma_{j}$ is the uncertainty per bin.

\begin{figure}[t]
\includegraphics[width=0.43\columnwidth]{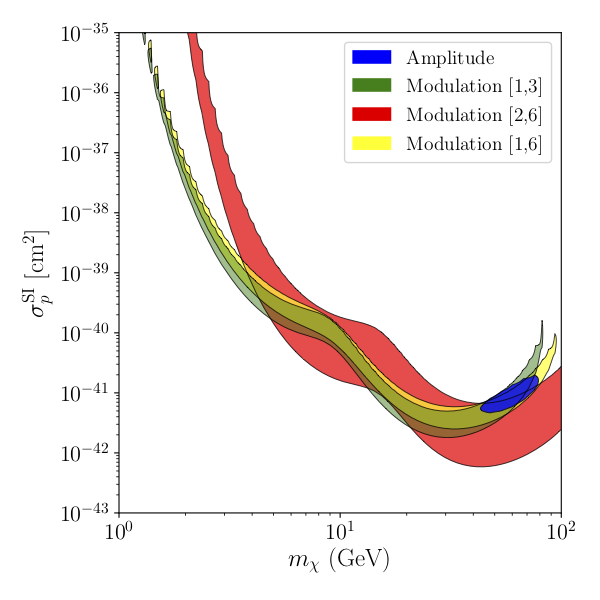}
\includegraphics[width=0.43\columnwidth]{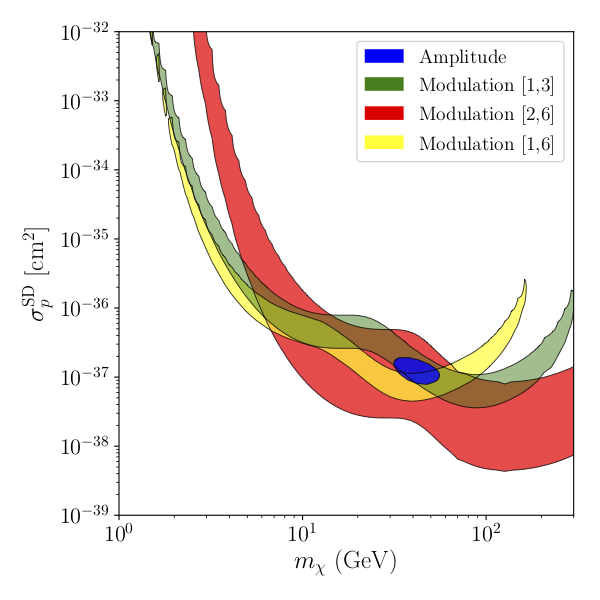}
\caption{Best fit $2\sigma$ regions for the DAMA/Libra phase-2 data \cite{Bernabei:2018yyw}. The blue region is the best-fit to the reported modulation amplitude as a function of electron recoil energy, the green, red, and yellow regions are the fits to the annual modulation data in the $[1,3]$, $[2,6]$, and $[1,6]$~keV$_{\rm ee}$ recoil energy bins respectively. Left: $2\sigma$ spin-independent limits and right: $2\sigma$ spin-dependent limits. All fits assume the SHM velocity distribution and isospin-conserving couplings. \label{fig:DAMArecast} }
\end{figure}

For the modulation amplitude we assume a period of one year for the modulation, $\omega = 2\pi/(365.25~{\rm days})$, and extract the amplitude of this mode from Eq.~\eqref{eq:electron_equiv} per energy bin:
\begin{equation}
S_{m,k} (m_{\chi}, \sigma_{p}) =  \frac{2}{\Delta E_e\times (365.25~{\rm days}) }\int_{E_{e,k}^{\rm min}}^{E_{e,k}^{\rm max}} \int_{0}^{365.25~\rm days} \cos(\omega t )\frac{dR}{dE_{e}}(m_{\chi}, \sigma_{p}) ~dE_{e} ~dt.
\end{equation}
Using a similar form as Eq.~\eqref{eq:chi2res} above, we perform a chi-squared fit to the \dama~phase-2 amplitude data \cite{Bernabei:2018yyw} binned in $0.5$~keV bins from $E_e = 1-4$~keV, 1~keV bins from $4-7$~keV, and one bin from $6-20$~keV. We fit the mass $m_\chi$ and dark matter-proton scattering cross section $\sigma_p$, assuming a SHM velocity distribution and isospin-conserving interaction. 
The resulting best-fit regions as a function of $m_\chi$ and $\sigma_p^{\rm SI}$ are shown in Figure~\ref{fig:DAMArecast}, (spin-independent on the left and spin-dependent on the right) and broadly match the results of Ref.~\cite{Baum:2018ekm}. Both amplitude and annual modulation confidence regions are overlaid in Fig.~\ref{fig:DAMArecast} with the 1-3 keV, 1-6 keV, and 2-6 keV energy range modulation fits represented by the green, red and yellow shaded contours respectively and the amplitude data represented by the blue shaded region.

\subsection{Xenon-1T}
\begin{figure}[t]
\includegraphics[width=0.4\columnwidth]{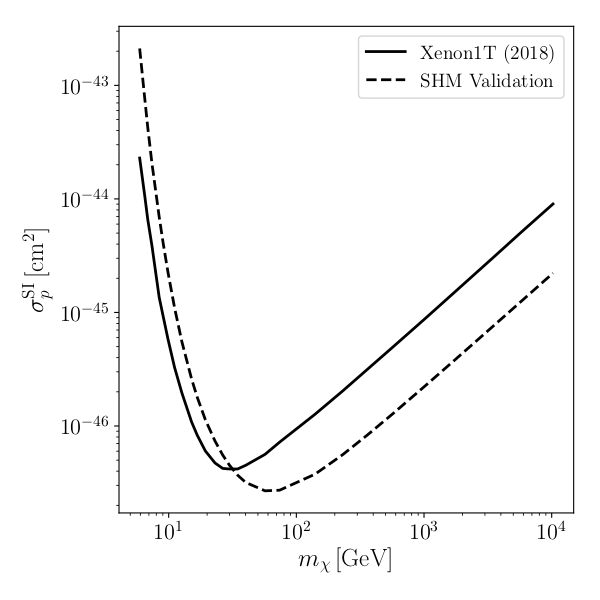}
\hspace*{0.3cm}
\includegraphics[width=0.4\columnwidth]{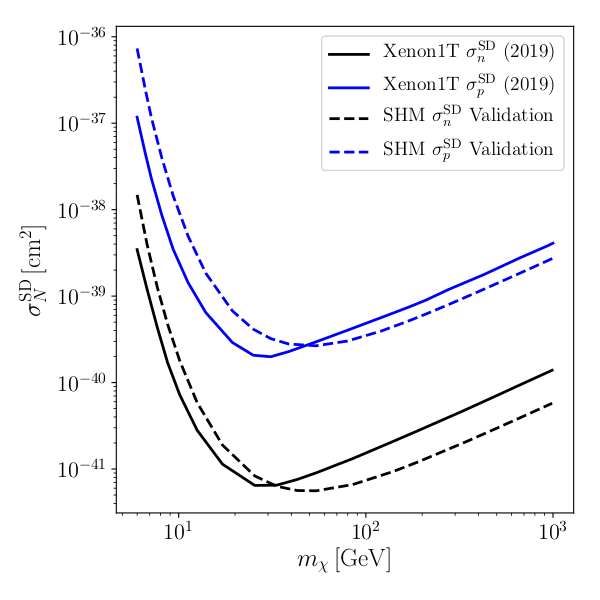}
\caption{Left: 90\% CL upper limits (solid black) on isospin-conserving spin-independent proton cross section $\sigma_p^{\rm SI}$ set by \xenon~\cite{Aprile:2018dbl} with our validation at 90\% CL shown as the black dashed line. Right: 90\% CL upper limits on the isospin conserving spin-dependent proton (blue solid) and neutron (black solid) cross-sections with our validation represented as the dashed lines. The y-axis label $\sigma_{N}^{SD}$ is intended to represent the respective nucleon. Our validations both assume the Standard Halo Model velocity distribution. 
\label{fig:Xenon1TRecast} }
\end{figure}

\xenon~is a liquid xenon detector which is sensitive to the scintillation light of dark matter-nucleon interactions. With a total mass of 1.3 tonnes and 280 days of exposure, \xenon~sets, at the time of writing, the strongest limits on dark matter-nucleon spin-independent scattering. We adopt the limits from Ref.~\cite{Aprile:2018dbl} for spin-independent scattering and from \cite{Aprile:2019dbj} for spin dependent scattering. \xenon, like most liquid xenon-based detectors, uses two scintillation signals ($S1$ and $S2$) to reject backgrounds. 
We use the published efficiency curves from Refs.~\cite{Aprile:2018dbl, Aprile:2019dbj} for validation of our calculations. 
In the \xenon~signal region we set our 90\% CL upper limit (solid black) to correspond to cross sections that give 3.7 events. Our SHM validation curves are shown in Figure~\ref{fig:Xenon1TRecast} (on the left are spin-independent (dashed black) and on the right are the spin-dependent limits for dark matter - proton (dashed blue) and - neutron (dashed black) interactions).  As can be seen, at low recoil energy (corresponding to low dark matter mass) it is difficult to fully model the detector response, which is rapidly changing in this regime, resulting in recast limits that are off by ${\cal O}(1)$ from the official collaboration results.

\subsection{CDMSlite}
\cdms~is part of the \textsc{SuperCDMS} experiment \cite{Agnese:2013jaa,Agnese:2015nto}, run in a low-threshold mode to maximize sensitivity to low mass ($\lesssim 10$~GeV) dark matter. It uses germanium targets that measure both ionization and phonons to search for dark matter recoils, rejecting background using the ratio of ionization to phonon energy (the ionization yield). 

\begin{figure}[t]
\includegraphics[width=0.45\columnwidth]{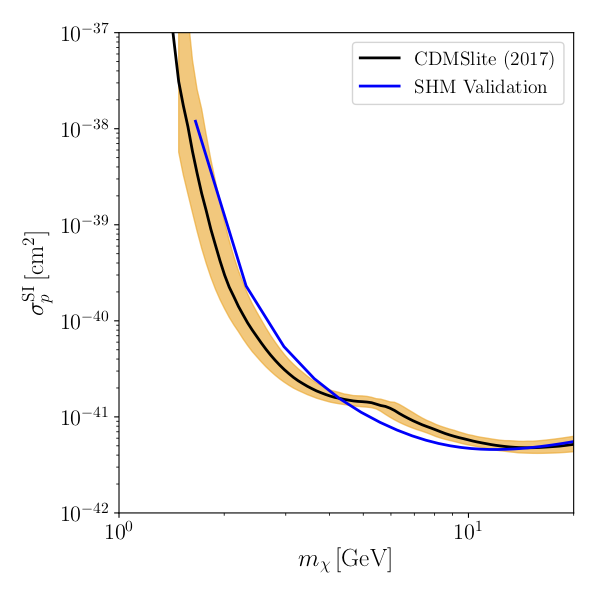}
\caption{Observed (solid black) 90\% CL upper limits on isospin-conserving spin-independent proton cross section $\sigma_p^{\rm SI}$ set by \cdms~Run-2 \cite{Agnese:2017jvy} along with the 95\% confidence region. Our validation 90\% CL upper bound is shown in blue assuming the Standard Halo Model velocity distribution. \label{fig:CDMSliteRecast} }
\end{figure}

We adopt the limits set by \cdms~in Ref.~\cite{Agnese:2017jvy}, using that reference's parameters for the Lindhard model for the relation between electron equivalent and nuclear recoil energies. We recast the limits from \cdms's Run 2, using the Ref.~\cite{Agnese:2017jvy} reported energy-dependent efficiency and an exposure of 70.1~kg$\times$day. We set our recast limits using the published background spectrum and the maximum gap method \cite{Yellin:2002xd} in the energy range of $[0.3,1]$~keV$_{\rm ee}$. The comparison between our recast limits assuming the SHM and the \cdms~results are shown in Figure~\ref{fig:CDMSliteRecast} as the solid blue line. 

\subsection{PICO-60}
\pico~is a superheated bubble chamber filled with 52.2~kg of C$_3$F$_8$ target. Nuclear recoils that deposit energy above the $3.3$~keV threshold cause a bubble nucleation. The acoustic properties of these bubbles can be used to discriminate between alpha decays and the dark matter-signal nuclear recoils. Due to the nuclear high spin of fluorine, \pico~provides the strongest constraints on spin-dependent, dark matter-proton interactions. We use the results of Ref.~\cite{Amole:2017dex}, corresponding to 1167~kg$\times$days of exposure. 

We use the bubble nucleation efficiency curve for fluorine as reported by \pico~in Ref.~\cite{Amole:2015lsj}. With an exposure of 1167~kg$\times$days \pico~observed zero single bubble events, and expected a background of 0.25 events. Assuming a Poisson distribution, the 90\% C.L. upper limit on the number of signal events is 2.05. Our validation of the upper limit is shown in the blue solid line with the \pico~reported limit shown in black. Our validation assumes the SHM velocity distribution and we show only the spin-dependent limits since they are currently the most competitive.

\begin{figure}[t]
\includegraphics[width=0.45\columnwidth]{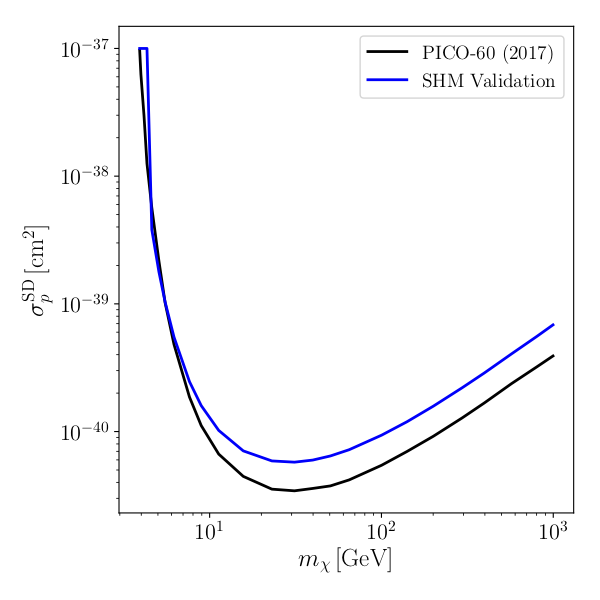}
\caption{Observed (solid black) 90\% CL upper limits on the spin-dependent proton cross section $\sigma_p^{\rm SD}$ set by \pico~\cite{Amole:2017dex}. Our validation 90\% CL upper bound is shown in blue assuming the Standard Halo Model velocity distribution. 
\label{fig:PICO60Recast} }
\end{figure}

\subsection{Cosine-100}
\cosine~is a direct detection experiment using NaI crystal targets, recording nuclear recoil signals using the emitted light, as measured by photomultiplier tubes. This experiment, using the same target as \dama, is an attempt to directly measure that experiment's claimed signal in an identical material. At present, the sensitivity of \cosine~is not sufficient to see the claimed modulation signal from \dama, but a limit can be set on the dark matter scattering rate averaged over the detector live-time (59.5 days between Oct.~20 and Dec.~19, 2016). Total target mass is 105~kg, though only $79$~kg was used in the analysis~\cite{Adhikari2018}. 

\begin{figure}[t]
\includegraphics[width=0.45\columnwidth]{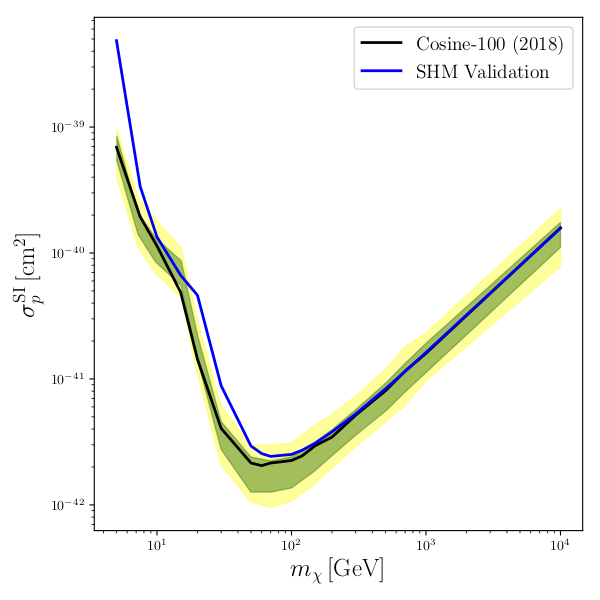}
\caption{Observed (solid black) 90\% CL upper limits on isospin-conserving spin-independent proton cross section $\sigma_p^{\rm SI}$ set by Cosine-100 \cite{Adhikari2018}. along with the 1 and $2\sigma$ confidence regions. Our validation 90\% CL upper bound is shown in blue assuming the Standard Halo Model velocity distribution. \label{fig:CosineRecast} }
\end{figure}

We use the efficiency curve provided in Ref.~\cite{Adhikari2018}, along with their background model in the recoil energy range $2-6$~keV. Using the provided event spectrum, background rate, and estimated errors, we set a 90\% CL limit on the scattering cross section using a simple $\chi^2$ measure. The comparison between our estimate for the limits on cross section assuming the SHM model and the results of Ref.~\cite{Adhikari2018} are shown in Figure~\ref{fig:CosineRecast}. Our validation is shown in blue, while the reported \cosine~limit is shown in black with $1$ and $2\sigma$ uncertainties. We note here that another independent experiment in Spain called \anais~\cite{Amare:2019jul} released results, running with roughly similar exposure (157 kg.y) to \cosine. Both experiments do not see any large evidence of annual modulation and expect to cover the \dama~region at 3$\sigma$ only in 5 years. Since their modulation phase results are roughly similar and they only differ mildly in the amplitude fits, including \anais~ in our analysis is not expected to cover different parameter space not already covered by \cosine. Hence we do not include \anais~in our analysis, but cite it here.  

\bigskip

\bibliography{gaiadama}
\end{document}